\documentclass[12pt]{article}

\usepackage{PRIMEarxiv}

\usepackage{graphicx}

\usepackage{diagbox}
\usepackage[subrefformat=parens]{subcaption}

\usepackage{float}

\usepackage{amsmath}
\usepackage{amssymb}
\usepackage{hyperref}
\usepackage{physics}
\usepackage{dsfont}

\usepackage{tablefootnote}
\usepackage{makecell}

\usepackage{algorithm, algorithmic}

\usepackage[backend=bibtex]{biblatex}
\newcommand{\R}{\mathbb{R}}

\allowdisplaybreaks

\title{VISTA-SSM: Varying and Irregular Sampling Time-series Analysis via State Space Models}

\author{
\textbf{Benjamin Brindle}$^1$ \quad \textbf{Thomas Derrick Hull}$^2$ \quad \textbf{Matteo Malgaroli\dag}$^{3}$ \quad \textbf{Nicolas Charon\dag}$^{4}$\\ \\
$^1$ Department of Applied Mathematics and Statistics, Johns Hopkins University\\
$^2$ Institute for Psycholinguistics and Digital Health, United States of America\\
$^3$ Department of Psychiatry, New York University School of Medicine\\
$^4$ Department of Mathematics, University of Houston \quad
$^\dag$ Equal contribution\\ \\
\textbf{Correspondence:} Matteo Malgaroli, PhD. One Park Avenue, 8th Floor, New York, NY 10016, USA.\\ Phone: 646-754-4030. Email: matteo.malgaroli@nyumc.org
}

\begin{document}
\maketitle
\begin{abstract}
    We introduce VISTA, a clustering approach for multivariate and irregularly sampled time series based on a parametric state space mixture model. VISTA is specifically designed for the unsupervised identification of groups in datasets originating from healthcare and psychology where such sampling issues are commonplace. Our approach adapts linear Gaussian state space models (LGSSMs) to provide a flexible parametric framework for fitting a wide range of time series dynamics. The clustering approach itself is based on the assumption that the population can be represented as a mixture of a fixed number of LGSSMs. VISTA's model formulation allows for an explicit derivation of the log-likelihood function, from which we develop an expectation-maximization scheme for fitting model parameters to the observed data samples. Our algorithmic implementation is designed to handle populations of multivariate time series that can exhibit large changes in sampling rate as well as irregular sampling. We evaluate the versatility and accuracy of our approach on simulated and real-world datasets, including demographic trends, wearable sensor data, epidemiological time series, and ecological momentary assessments. Our results indicate that VISTA outperforms most comparable standard times series clustering methods. We provide an open-source implementation of VISTA in Python.
\end{abstract}


\section{Introduction}
Advancements in digital measurement have enabled more consistent and comprehensive tracking of outcomes, treatment adherence, and other relevant metrics in psychological and healthcare sciences \parencite{galatzer2023machine,wang2018bigdata,russell2020annual}. These improvements have been driven by innovations in mobile applications, wearable technology, and electronic health records (EHRs) \parencite{adler2017HITECH}, which facilitate the collection of vast amounts of data from diverse sources \parencite{bates2014bigdata}. For example, integrating wearable devices into routine healthcare allows continuous monitoring of physiological parameters free from the distortions inherent in self-reported data \parencite{piwek2016rise}, and providing a more comprehensive picture of patient health \parencite{topol2019highperf}. 

Despite the potential of digital and time-intensive data, three significant barriers hinder its effective use in real-world settings. The first barrier is capturing high-quality data in under-resourced environments \parencite{biber2013challenge} lacking infrastructure due to associated costs, technical expertise, and logistical challenges \parencite{raghupathi2014bigdata}. Neuroimaging data, including EEG, might be captured at varying intervals depending on patient compliance, experimental design, or technical issues, leading to datasets with non-uniform temporal resolutions \parencite{poldrack2015long}. This barrier is exacerbated in para-clinical settings, such as community health programs or primary care, where even basic data collection can be inconsistent and fragmented \parencite{frenk2010democ}. As a result, many interventions are analyzed using incomplete data \parencite{pratap2020indicators}, which may not capture the complexities of the intervention itself. The second barrier is that data collected in psychological sciences are often indirect representations of the phenomena of interest. For instance, patient-reported outcomes in healthcare may be affected by mood or recall accuracy \parencite{stone2002capturing}, while macro indicators often fail to fully capture granular real-time dynamics \parencite{stiglitz2009econ}. These challenges are especially pronounced when attempting to identify clinically relevant subgroups \parencite{insel2014NIMH}, or when evaluating interventions with mechanisms that are not yet fully understood \parencite{hofman2012efficacy}. The third barrier is that clinical data is inherently characterized by irregular response intervals and events. This issue is particularly evident in contexts where data capture is continuous or semi-continuous, such as mental health interventions \parencite{jung2021deeplearning}. In psychotherapy, in particular, the content and timing of interactions are highly variable, reflecting the dynamic and responsive nature of therapeutic conversations \parencite{malg2023nlpreview}. Capturing and analyzing such data in real-world settings, where consistency and control are often limited, presents a unique challenge requiring specialized methods capable of handling irregularities in the data structure.

Traditional methods assume regularity in data collection, making them ill-suited for handling the complexities of real-world data that are characterized by irregular sampling, missing data points, or variable temporal structures \parencite{kreindler2016effects}. Analyzing data with irregular or ill-structured intervals necessitates new methods that address these challenges while also leveraging the irregularity as a potential source of insight into underlying processes. There is a pressing need to develop new methods that can not only handle these sampling issues but also use them to help identify relevant subgroups within the dataset that exhibit distinct dynamical patterns in their time series.

Accordingly, we introduce a new approach for handling Varying and Irregular Sampling Time-series Analysis (VISTA), specialized in the automatic and unsupervised clustering of subgroups that exhibit distinct dynamic patterns. This is a common research challenge in psychological sciences when exploring temporal data without clear prior categorical information about the subjects in a population, e.g. \parencite{bonanno2023resilience}. We conduct extensive evaluation of VISTA on datasets from various sources and dimensionality, illustrating the versatility of the approach in finding meaningful subgroups even without necessarily requiring prior knowledge of the number of clusters. 

\subsection{Related work}
There has been considerable interest and effort in developing clustering methods in particular for time series. These can be broadly assigned to three categories; c.f. \parencite{liao2005clustering}, 1) \textit{raw-data based} approaches that directly operate on a certain similarity measure between time series, such as the dynamic time warping or elastic distance \parencite{srivastava2011registration} 2) \textit{feature-based} methods that first extracts a feature representation of the data, e.g. wavelet decomposition \parencite{keogh2002wavelet} or more recently latent representation estimated by training neural networks \parencite{ma2019learning} and 3) \textit{model-based} approaches which attempt to fit a lower-dimensional predefined parametric model to the observed data. In healthcare and psychology settings, the latter model-based approach has so far been more commonly used. This is due, in part, to the aforementioned specific data challenges, but also to the fact that they usually lead to more interpretable results while also offering forecasting, prediction, or missing data imputation based on the fitted model. Among the most usual families of parametric frameworks for time series are the hidden Markov models (HMM) and the \textit{autoregressive models} described in e.g. \parencite{hamilton2020time}. With HMM being mostly applicable to discrete latent variables, many studies have instead relied on AR/ARMA models. These include a number of recent works relevant to psychology such as \parencite{jebb2015time,bulteel2016clustering,haslbeck2023testing,park2024subgrouping}. Yet a well-known limitation of AR and ARMA is the fact that they are not well-adapted to the modeling of non-stationary processes thus negatively impacting their performance when dealing with time series data exhibiting time varying expectation or covariance. Despite these limitations, recent work relevant to psychology has successfully modified the AR framework to tackle the challenging types of data described above, with \parencite{ernst2020inter,liu2021dynamic} using discrete-time models and \parencite{driver2017continuous,driver2018hierarchical,ou2019s} continuous-time models.

Those issues have led researchers to consider alternative parametric frameworks such as \textit{Linear Gaussian State Space models} (or LGSSMs) originating from the field of control theory \parencite{ho1964bayesian}. In this model, each observed time series is viewed as a noisy projection of a continuous higher-dimensional latent state variable with its evolution described by a linear stochastic equation. This has the advantage of being considerably more flexible than the AR model in representing more complex dynamical patterns while still leading to simple expressions for the probability distributions of both latent and observed variables, and thus to a tractable closed form for the likelihood function through Kalman filtering \parencite{kalman1961new,kitagawa1996smoothness}. Furthermore, LGSSMs can be extended to tackle time series clustering by introducing a mixture of LGSSMs and jointly estimating the cluster assignments together with the model parameters, as was recently proposed in \parencite{umatani2023time}. This work demonstrated clear performance improvement over the aforementioned approaches, although validation is mainly restricted to small one-dimensional time series datasets. There are still several obstacles in successfully adapting this method to the more challenging types of data described earlier. First among these is the assumption of consistent time sampling across different time series in the dataset leading to a numerical implementation specifically tailored to that situation. 

\subsection{Contributions}
\begin{figure}
    \centering
    \includegraphics[width=1.0\linewidth]{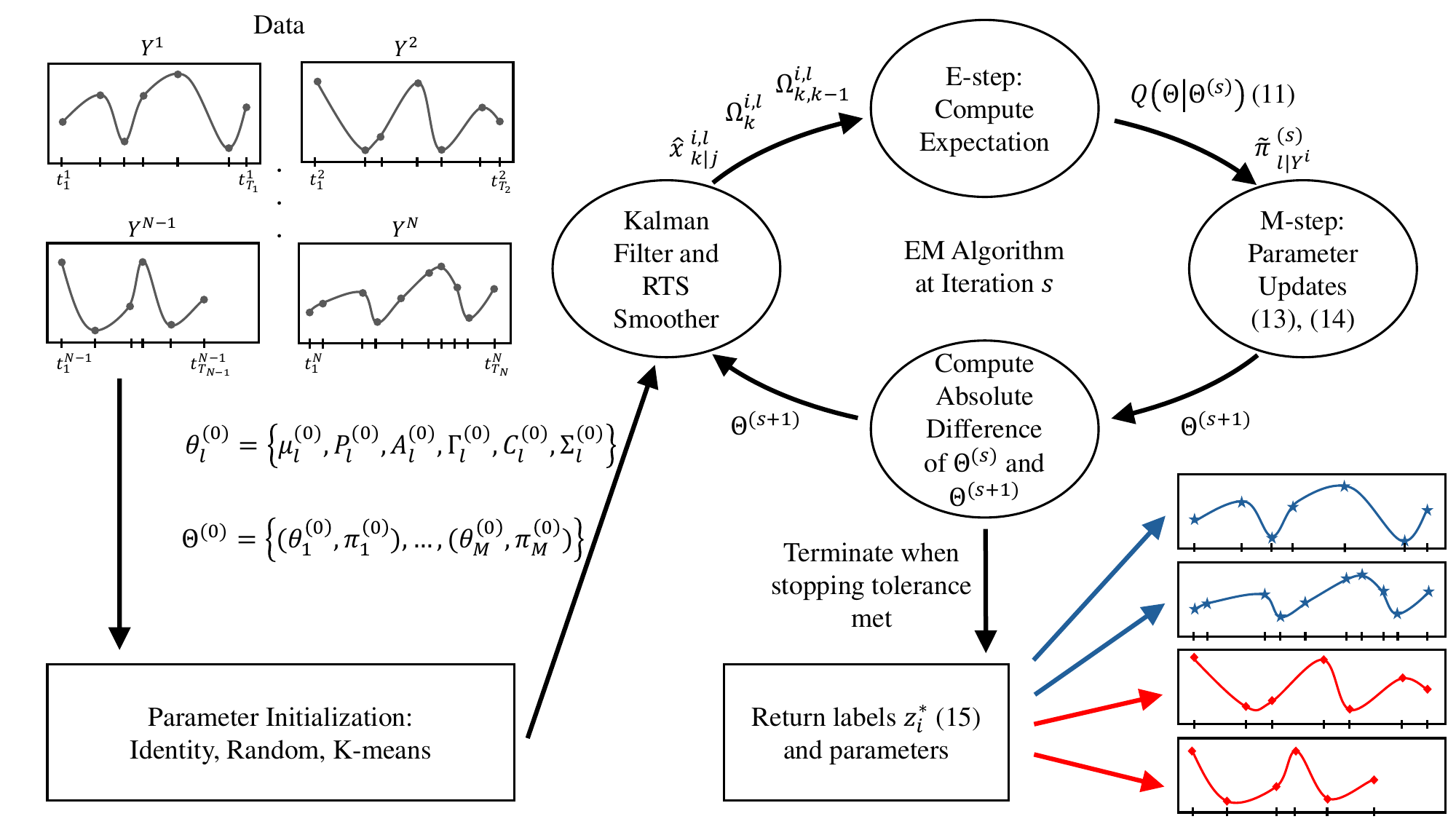}
    \caption{Schematic of our algorithm.}
    \label{fig:schematic}
\end{figure}

One of the core contributions of VISTA lies in its generalized formulation of LGSSMs and mixture of LGSSMs for time series data with irregular and inconsistent sampling. We emphasize that this is fundamentally different than dealing with missing samples in otherwise consistently sampled data, as this type of situation can be directly tackled by standard parametric models like AR or LGSSM based on forecasting. In contrast, VISTA leverages an underlying continuous stochastic process formulation to derive a consistent LGSSM for irregular time steps. To make the presentation self-contained, we further show how the estimation of the latent variables can be reduced to the usual Kalman filtering and smoothing scheme. We then introduce the corresponding mixture of LGSSMs from which we derive our clustering approach based on the expectation-maximization algorithm. Our Python implementation is able to handle time series with wide variations in sampling rate while preventing memory overflow errors when processing large cohorts with potentially high temporal resolution. The entire process is shown in schematic form in Figure \ref{fig:schematic}. To allow for reproducibility and provide a potentially valuable analysis tool, all codes and examples of this paper are made available at \url{https://github.com/benjaminbrindle/vista_ssm}.

\section{Linear Gaussian State Space Models}
\label{sec:single_LGSSM}
We start by introducing the linear Gaussian state space model for a single time series, which we shall then extend to the mixture scenario in the next section.  

\subsection{Dynamical system}
In order to define a LGSSM for irregularly sampled time series, let us first consider the following continuous time-invariant linear stochastic dynamical system:
\begin{equation}\label{eq:continuous}
    \begin{split}
        \dot x(t) &= Ax(t) + w(t) \\
        y(t) &= C x(t) + v(t)
    \end{split}
\end{equation}
in which, for any given time $t$, $y(t) \in \R^n$ stands for the observed variable of the system while $x(t) \in \R^d$ represents a latent, unobserved (a.k.a state) variable, with typically $d \geq n$.  In the language of state-space models, $A \in \R^{d \times d}$ may be interpreted as the system matrix which describes the underlying linear dynamics of the latent variables while $C \in \R^{n \times d}$ is the output matrix relating the latent to the observed variables. The noise variables $w(t)$ and $v(t)$ are modeled as white noise processes on $\R^d$ and $\R^n$ respectively, following centered Gaussian laws $w(t) \sim N(0,\Gamma)$, and $v(t) \sim N(0,\Sigma)$, $\Gamma$ and $\Sigma$ being symmetric positive definite matrices of size $(d \times d)$ and $(n \times n)$ respectively. In addition, the initial value of the latent variable is assumed to be also random, such that $x(0) = \mu + u$ with $u \sim N(0,P)$ and $\mu\in \R^d$ represents the initial drift of $x$. It is further assumed that all the different noise variables are non-correlated. As we explain next, the LGSSM that we will be focusing on in the rest of this paper can be properly derived as a time discretization of the above system.

Let $T$ be a positive integer and $\{t_k\}_{k=1,\ldots,T}$ a discrete sequence of time stamps with $t_{k-1} < t_{k}$ for all $k=2,\ldots,T$ and we assume, without loss of generality, that $t_1=0$. We shall write $\{y(t_k)\}_{k=1}^{T}$ as the corresponding sequence of observed variables and $\{x(t_k)\}_{k=1}^{T}$ as the sequence of latent variables at those same time points. Let us fix a specific $k\in\{2,\ldots,T\}$ and denote by $\Delta_k = t_{k} - t_{k-1}$. Integrating \eqref{eq:continuous} between $t_{k-1}$ and $t_{k}$ leads to:
\begin{equation*}
x(t_{k}) = e^{\Delta_k A} x(t_{k-1}) + \int_{t_{k-1}}^{t_{k}} e^{(t_{k}-t)A} w(t) dt
\end{equation*}
So as to recover a tractable model for the later estimation of the parameters, we first make the small time step first-order approximation $e^{\Delta_k A} \approx \text{Id} + \Delta_k A$. Similarly, defining $w_k = \int_{t_{k-1}}^{t_{k}} e^{(t_{k}-t)A} w(t) dt$, we have:
\begin{equation*}
 w_k = \int_{0}^{\Delta_k} e^{sA} w(t_{k}-s) ds
\end{equation*}
Using the properties of the continuous white noise process $w(t)$, one can show that $w_k$ is a Gaussian random variable of zero mean with covariance matrix:
\begin{equation*}
 \Gamma_k = \int_0^{\Delta_k} e^{tA} \Gamma e^{tA^T} dt
\end{equation*}
for which we again use a first-order approximation with respect to the time step to obtain $\Gamma_k \approx \Delta_k \Gamma$. The second equation in \eqref{eq:continuous} simply expresses the observed variable as a linear ``projection'' of the latent one with additional noise at each time $t$. As such, it is naturally discretized as $y(t_k) = C x(t_k) + v_k$. Here, $v_k$ is Gaussian random noise with zero mean. Its covariance matrix should be related to $\Sigma$ so that the discrete equations approximate the continuous process. As argued in \parencite{lewis2017optimal}, this entails that $v_k \sim N(0,\Sigma_k)$ with $\Sigma_k = \Sigma/\Delta_k$.

Summarizing the above, let us therefore introduce the two sequences $X=\{x_k\}_{k=1}^{T}$ and $Y=\{y_k\}_{k=1}^{T}$ corresponding to the latent and observed variables at the $t_k$'s, and consider the following stochastic dynamical system:
\begin{equation}\label{eq:lgssm}
    \begin{split}
        x_k &= A_k x_{k-1} + w_k\\
        y_k &= C x_k + v_k\\
        x_1 &= \mu + u
    \end{split}
\end{equation}

in which $A_k = \text{Id} + \Delta_k A$ and $\{w_k\}$, $\{v_k\}$, and $u$ are Gaussian i.i.d random variables with zero mean and respective covariance matrices $\Gamma_k = \Delta_k \Gamma$, $\Sigma_k = \frac{1}{\Delta_k}\Sigma$ and $P$, respectively.

Now, assuming that a time series of data points $\{y_k\}_{k=1}^T$ is observed at the times $t_k$ and that it follows that the stochastic system prescribed by \eqref{eq:lgssm} for a given dimension $n$ but unknown set of parameters $\theta=\{\mu,A,C,\Gamma,\Sigma,P\}$, the question we wish to address is how to estimate those parameters. Due to the presence of the unknown latent variables $\{x_k\}$ in the model, standard maximum likelihood estimation is not straightforward and one common approach is to instead rely on an Expectation-Maximization (EM) strategy.

\subsection{Latent variable inference}
\label{ssec:latent_inference}
In order to derive the estimation algorithm, let us first describe how the latent variables can be inferred from the observations, assuming for now that the model parameters $\theta$ are fixed. This can be done with classical Kalman filtering \parencite{kalman1961new} and Rauch-Tung-Striebel smoothing  \parencite{rauch1965maximum}. For completeness, we briefly recap the equations and algorithm associated with the class of LGSSMs described by \eqref{eq:lgssm}, we refer the reader to \parencite{hamilton2020time} Chap. 13 or \parencite{anderson1979optimal} Chap. 3 for details. 

A remarkable and very convenient property of LGSSMs is that all observed and latent variables can be shown to be Gaussian distributed. Thus, the quantities to estimate are the following state conditional expectation and covariance which we will denote: $\hat{x}_{k|T} = E(x_k|Y,\theta)$ and $P_{k|T} = E((x_k-\hat{x}_{k|T})(x_k-\hat{x}_{k|T})^T|Y,\theta)$. This estimation is typically done in two successive stages. Let $j\leq k$ and introduce the notations $Y_{j}=\{y_k\}_{k=1}^{j}$ and $\hat{x}_{k|j} = E(x_k|Y_{j},\theta)$, $P_{k|j} = E((x_k-\hat{x}_{k|j})(x_k-\hat{x}_{k|j})^T|Y_{j},\theta)$. 

In the \textit{filtering} stage, one recursively computes the \textit{a priori} state and covariance estimates $\hat{x}_{k|k-1}$ and $\Gamma_{k|k-1}$ with the initializations $\hat{x}_{1|0} = E(x_1)=\mu$ and $P_{1|0} =P$ and for $k=2,\ldots,T$: 
\begin{equation}
\label{eq:Kalman_filtering1}
\left\{\begin{aligned}
    &\hat{x}_{k|k-1} = A_{k} \hat{x}_{k-1|k-1}  \\
    &P_{k|k-1} = A_k P_{k-1|k-1} A_k^T + \Gamma_{k} 
\end{aligned}
\right.
\end{equation}
together with the \textit{a posteriori} estimates $\hat{x}_{k|k}$ and $\Gamma_{k|k}$ which are obtained from the following iterative equations for $k=1,\ldots,T$:
\begin{equation}
\label{eq:Kalman_filtering2}
\left\{\begin{aligned}
    &K_k = P_{k|k-1} C^T(CP_{k|k-1}C^T + \Sigma_k)^{-1} \\
    &\hat{x}_{k|k} = \hat{x}_{k|k-1} + K_k(y_k-C\hat{x}_{k|k-1}) \\
    &P_{k|k} = (\text{Id} - K_kC)P_{k|k-1}
\end{aligned} \right.
\end{equation}
The matrix $K_k$ above is known as the Kalman gain matrix. The combined equations \eqref{eq:Kalman_filtering1} and \eqref{eq:Kalman_filtering2} lead to a forward iterative approach for jointly computing $\hat{x}_{k|k-1}$, $P_{k|k-1}$, $\hat{x}_{k|k}$ and $P_{k|k}$.

Once the previous a priori and a posteriori estimates have been obtained, the state's expected values and covariances conditioned on the whole set of observations $Y=Y_T$ are recovered via a \textit{smoothing} step. We consider here the Rauch-Tung-Striebel smoother. It amounts to iterating from $T-1$ backwards to $k=1$ the following:
\begin{equation}
  \label{eq:Kalman_smoothing}
\left\{\begin{aligned}
    &J_k = P_{k|k} A_{k+1}^T P_{k+1|k}^{-1} \\
    &\hat{x}_{k|T} = \hat{x}_{k|k} + J_k(\hat{x}_{k+1|T}-\hat{x}_{k+1|k}) \\
    &P_{k|T} = P_{k|k} + J_k(P_{k+1|T}-P_{k+1|k})J_k^T
\end{aligned} \right.  
\end{equation}
In conclusion, for each $k=1,\ldots,T$, we have $x_k|Y,\theta \sim N(\hat{x}_{k|T},P_{k|T})$ with $\hat{x}_{k|T},P_{k|T}$ obtained via the above filtering and smoothing scheme. In other words, the posterior distribution for the latent variable $X$ is:
\begin{equation*}
   X|Y,\theta \sim \prod_{k=1}^{T} N(\hat{x}_{k|T},P_{k|T}). 
\end{equation*}
We also compute the probability of an observation $Y$ given the parameters $\theta$ as follows:
\begin{equation*}
    p(Y|\theta) = p(y_1|\theta) \, p(y_2|y_1,\theta) \, \ldots \, p(y_T|Y_{T-1},\theta)
\end{equation*}
and for each $k=1,\ldots,T$, we have $y_k = C x_k + v_k$ and, following the above results, $x_k|Y_{k-1},\theta \sim N(\hat{x}_{k|k-1},\Gamma_{k|k-1})$ from which we get that $y_k|Y_{k-1},\theta \sim N(C\hat{x}_{k|k-1},C\Gamma_{k|k-1}C^T + \Sigma_k)$. Therefore, we have:
\begin{equation}
\label{eq:prob_Y_theta}
    Y|\theta \sim \prod_{k=1}^{T} N(C\hat{x}_{k|k-1},C\Gamma_{k|k-1}C^T + \Sigma_k).
\end{equation}

\subsection{EM parameter estimation approach}
\label{ssec:EM_LGSSM}
\subsubsection{Complete data log-likelihood}
Let us now return to the estimation of the LGSSM parameters $\theta$ from the observation of a single time series $Y=\{y_k\}_{k=1}^{T}$. In the absence of a closed-form expression for the likelihood function, the idea is to rely on the expectation-maximization algorithm. The scheme itself first involves the computation of the complete data log-likelihood function for the model. For any given $X=\{x_k\}$ and $Y=\{y_k\}$, using the independence assumptions repeatedly, we get:
\begin{align*}
 p(X,Y \mid \theta) &= p(Y\mid X,\theta)\, p(X\mid \theta) \\
 &=\left(\prod_{k=1}^{T} p(y_k \, \big | \, x_k,\theta) \right) . p(x_1 \mid \theta). \left(\prod_{k=2}^{T} p(x_k \, \big | \, x_{k-1},\theta) \right)
\end{align*}
which, by taking the logarithm, leads to:
\begin{align*}
 \log \, p(X,Y \mid \theta) = &\sum_{k=1}^{T} \log \, p(y_k \, \big | \, x_k,C,\Sigma) \\
 &+ \log \, p(x_1 \mid \mu,P) + \sum_{k=2}^{T} \log \, p(x_k \, \big | \, x_{k-1},A,\Gamma).
\end{align*}
Now, based on \eqref{eq:lgssm} and the probability laws for the noise variables $w$, $v$ and $u$, we obtain the following complete data log-likelihood function for our model:
\begin{align}
\label{eq:CDLL_LGSSM}
 &\log \, p(X,Y \mid \theta) = \nonumber\\
 &-\sum_{k=1}^{T}\frac{\Delta_k}{2} (y_k - C x_k)^T \Sigma^{-1} (y_k - C x_k) -   \frac{n}{2} \log(2\pi) - \frac{1}{2}\log |\Sigma| + \frac{1}{2}\log \Delta_k \nonumber\\
 &- \frac{1}{2} (x_1 - \mu)^T P^{-1} (x_1-\mu) - \frac{d}{2} \log(2\pi) - \frac{1}{2}\log |P| \nonumber\\
 &- \sum_{k=2}^{T} \frac{(x_{k} - A_k x_{k-1})^T \Gamma^{-1} (x_{k} - A_k x_{k-1})}{2\Delta_k} -  \frac{d}{2} \log(2\pi) - \frac{1}{2} \log |\Gamma| - \frac{1}{2}\log \Delta_k
\end{align}
where we have used the notation $|\cdot|$ for the determinant of a matrix. Omitting the constant terms (i.e. those not dependent on the model parameters), this can be rewritten more compactly as:
\begin{align}
 \label{eq:loglikeli_LGSSM}
 \log \, p(X,Y|\theta) =  &- \frac{T}{2} \log |\Sigma| - \frac{1}{2}\log |P| - \frac{T-1}{2} \log |\Gamma| - \frac{1}{2}(x_1 - \mu)^T P^{-1} (x_1-\mu) \nonumber \\
 &-\sum_{k=2}^{T} \frac{(x_{k} - A_k x_{k-1})^T \Gamma^{-1} (x_{k} - A_k x_{k-1})}{2\Delta_k} \nonumber\\
 &-\sum_{k=1}^{T}\frac{\Delta_k}{2} (y_k - C x_k)^T \Sigma^{-1} (y_k - C x_k) +\ldots
\end{align}

\subsubsection{E-step}
We may now derive the two steps of the EM algorithm for a single LGSSM, which is similar to the algorithm presented in \parencite{shumway1982approach,umatani2023time}. Denoting the current parameters' estimates at iteration $s$ by $\theta^{(s)}=\{\mu^{(s)},A^{(s)},C^{(s)},\mu^{(s)},\Gamma^{(s)},\Sigma^{(s)},P^{(s)}\}$, the E-step consists in computing the expectation of \eqref{eq:loglikeli_LGSSM} under the posterior distribution $p(X|Y,\theta^{(s)})$, namely:
\begin{align*}
    Q(\theta|\theta^{(s)}) &= E_{p(X|Y,\theta^{(s)})} \log \, p(X,Y \mid \theta) \\
    &=- \frac{T}{2} \log |\Sigma| - \frac{1}{2}\log |P| - \frac{T-1}{2} \log |\Gamma| -E((x_1 - \mu)^T P^{-1} (x_1-\mu)) \\
    &\phantom{=}-\sum_{k=2}^{T} \frac{1}{2\Delta_k} E((x_{k} - A_k x_{k-1})^T \Gamma^{-1} (x_{k} - A_k x_{k-1})) \\
    &\phantom{=}-\sum_{k=1}^{T}\frac{\Delta_k}{2} E((y_k - C x_k)^T \Sigma^{-1} (y_k - C x_k))  +\ldots
\end{align*}
Note that we used the notation $E(\cdot)$ in place of $E_{p(X|Y,\theta^{(s)})}(\cdot)$ for brevity. First, using the fact that $(x_1 - \mu)^T P^{-1} (x_1-\mu) = \text{Tr}((x_1 - \mu)^T P^{-1} (x_1-\mu))$, we get:
\begin{align*}
    &E((x_1 - \mu)^T P^{-1} (x_1-\mu)) = E(\text{Tr}(x_1^T P^{-1}x_1) -2\text{Tr}(x_1^T P^{-1}\mu) + \text{Tr}(\mu^T P^{-1}\mu)) \\
    &=\text{Tr}(P^{-1}E(x_1x_1^T)) - 2\text{Tr}(P^{-1}\mu E(x_1^T)) +\mu^T P^{-1}\mu
\end{align*}
Now, under the posterior $p(X|Y,\theta^{(s)})$, we have seen in the section on latent variable inference that $E(x_1) = \hat{x}_{1|T}$ and $E(x_1x_1^T) = P_{1|T} + \hat{x}_{1|T} \hat{x}_{1|T}^T$. Similarly, we write:
\begin{align*}
   &E((x_{k} - A_k x_{k-1})^T \Gamma^{-1} (x_{k} - A_k x_{k-1})) \\
   &=\text{Tr}(\Gamma^{-1}E(x_{k}x_{k}^T)) -2\text{Tr}(\Gamma^{-1}E(x_{k}x_{k-1}^T)A_k^T) + \text{Tr}(\Gamma^{-1}A_kE(x_{k-1}x_{k-1}^T)A_k^T)
\end{align*}
and, based again on the latent variable inference section, one has $E(x_{k}x_{k}^T) = P_{k|T} + \hat{x}_{k|T} \hat{x}_{k|T}^T$. The computation of $E(x_k x_{k-1}^T)$ is lengthier but can be adapted from the results of \parencite{shumway1982approach}. Similarly to the case of $E(x_{k}x_{k}^T)$, we write:
\begin{align*}
    &E(x_k x_{k-1}^T) = \text{Cov}_{p(X|Y,\theta^{(s)})}(x_k,x_{k-1}) + E(x_k)E(x_{k-1})^T = \text{Cov}_{p(X|Y,\theta^{(s)})}(x_k,x_{k-1}) + \hat{x}_{k|T}\hat{x}_{k-1|T}^T
\end{align*}
Furthermore, by law of total expectation:
\begin{align*}
    & \text{Cov}_{p(X|Y,\theta^{(s)})}(x_k,x_{k-1}) = E \qty((x_k - \hat x_{k|T})(x_{k-1}-\hat x_{k-1|T})^T) \\
    & = E \qty[E_{p(X|x_k,Y,\theta^{(s)})} \qty((x_k - \hat x_{k|T})(x_{k-1}-\hat x_{k-1|T})^T)] \\
    &= E \qty[(x_k - \hat x_{k|T})E_{p(X|x_k,Y,\theta^{(s)})} (x_{k-1}-\hat x_{k-1|T})^T]
\end{align*}
From \ref{eq:Kalman_filtering1}, \ref{eq:Kalman_filtering2}, and \ref{eq:Kalman_smoothing}, the equality follows:
\begin{align*}
    &E_{p(X|x_k,Y,\theta^{(s)})} (x_{k-1}-\hat x_{k-1|T}) = \hat x_{k-1|k-1} + J_{k-1}(x_k - \hat x_{k|k-1}) \\&- \hat x_{k-1|k-1} - J_{k-1}(\hat x_{k|T} - \hat x_{k|k-1}) = J_{k-1}(x_k - \hat x_{k|T})
\end{align*}
Returning to the covariance expression, this yields:
\begin{align*}
    & \text{Cov}_{p(X|Y,\theta^{(s)})}(x_k,x_{k-1}) = E \qty((x_k - \hat x_{k|T})(x_{k}-\hat x_{k|T})^T J_{k-1}^T) = P_{k|T}J_{k-1}^T
\end{align*}
This leads to $E(x_k x_{k-1}^T) = P_{k|T}J_{k-1}^T + \hat{x}_{k|T}\hat{x}_{k-1|T}^T$. Lastly, since the $y_k$'s are observed:
\begin{align*}
    &E((y_k - C x_k)^T \Sigma^{-1} (y_k - C x_k)) \\
    &=y_k^T \Sigma^{-1} y_k - 2\text{Tr}(\Sigma^{-1}y_k E(x_{k})^TC^T) + \text{Tr}(\Sigma^{-1}CE(x_{k}x_{k}^T)C^T) 
\end{align*}
To summarize, the E-step amounts to computing:
\begin{align}
\label{eq:ECDLL}
&Q(\theta|\theta^{(s)}) \nonumber\\ 
&= \frac{T}{2} \log |\Sigma^{-1}| + \frac{1}{2}\log |P^{-1}| + \frac{T-1}{2} \log |\Gamma^{-1}| - \frac{1}{2} \text{Tr}(P^{-1}\Omega_1) + \mu^TP^{-1}\hat{x}_{1|T} -\frac{1}{2} \mu^T P^{-1} \mu \nonumber \\
&\phantom{=}-\sum_{k=2}^{T} \frac{1}{2\Delta_k} \left(\text{Tr}(\Gamma^{-1}\Omega_k) - 2\text{Tr}(\Gamma^{-1}\Omega_{k,k-1}A_k^T) + \text{Tr}(\Gamma^{-1}A_k\Omega_{k-1}A_k^T)\right) \nonumber \\
&\phantom{=}-\sum_{k=1}^{T} \frac{\Delta_k}{2} \left(y_k^T \Sigma^{-1} y_k -2\hat{x}_{k|T}^TC^T\Sigma^{-1} y_k + \text{Tr}(\Sigma^{-1}C\Omega_k C^T)  \right) + \ldots
\end{align}
where $\Omega_k = E(x_{k}x_{k}^T) = P_{k|T} + \hat{x}_{k|T} \hat{x}_{k|T}^T$ and $\Omega_{k,k-1} =E(x_{k}x_{k-1}^T)=P_{k|T}J_{k-1}^T+\hat{x}_{k|T} \hat{x}_{k-1|T}^T$, with all the values of $\hat{x}_{k|T}$ and $P_{k|T}$ being computed based on the filtering and smoothing scheme associated to the current parameters $\theta^{(s)}$.

\subsubsection{M-step}
\label{ssec:M_step_LGSSM}
In the M-step, the parameters update $\theta^{(s+1)}$ is done by maximizing $Q(\theta|\theta^{(s)})$ in \eqref{eq:ECDLL} with respect to $\theta$. One can notice that the function is concave in $\mu,A$ and $C$ as well as in the inverse covariance matrices $\Sigma^{-1},\Gamma^{-1}$ and $P^{-1}$. We thus derive the update on $\theta$ by looking at the partial differentials of $Q(\theta|\theta^{(s)})$. Computations lead to the following:
\begin{align*}
  &\partial_{\mu} Q(\theta|\theta^{(s)})= -P^{-1}\mu + P^{-1}\hat{x}_{1|T}, \\  
  &\partial_{A} Q(\theta|\theta^{(s)}) = \sum_{k=2}^{T} \frac{1}{\Delta_k} \left(\Delta_k \Gamma^{-1} \Omega_{k,k-1} - \Delta_k \Gamma^{-1} A_k \Omega_{k-1} \right) \\
  &\partial_C Q(\theta|\theta^{(s)}) = \sum_{k=1}^{T} \Delta_k \Gamma^{-1}(y_k\hat{x}_{k|T}^T - C \Omega_k) \\
  &\partial_{P^{-1}} Q(\theta|\theta^{(s)})=\frac{P}{2} -\frac{\Omega_1}{2} +\frac{1}{2}(\hat{x}_{k|T}\mu^T + \mu \hat{x}_{k|T}^T - \mu \mu^T)\\
  &\partial_{\Sigma^{-1}}Q(\theta|\theta^{(s)}) = \frac{T}{2} \Sigma - \sum_{k=1}^{T} \frac{\Delta_k}{2}\left(y_ky_k^T- C \hat{x}_{k|T} y_k^T - y_k \hat{x}_{k|T}^T C^T + C \Omega_k C^T\right)\\
  &\partial_{\Gamma^{-1}}Q(\theta|\theta^{(s)}) = \frac{T-1}{2} \Gamma -\sum_{k=2}^{T}  \frac{1}{2\Delta_k} \left(\Omega_k-\Omega_{k,k-1}A_k^T-A_k\Omega_{k,k-1}^T+A_k\Omega_{k-1}A_k^T\right)
\end{align*}
The new set of parameters $\mu^{(s+1)}$, $A^{(s+1)}$, $C^{(s+1)}$, $P^{(s+1)}$, $\Sigma^{(s+1)}$ and $\Gamma^{(s+1)}$ is then obtained by following the sequence of updates below:
\begin{align}
\label{eq:LGSSM_Mstep}
&\mu^{(s+1)} = \hat{x}_{1|T}, \ \ A^{(s+1)}= \left(\sum_{k=2}^{T} \Omega_{k,k-1} - \Omega_{k-1} \right) \left(\sum_{k=2}^{T} \Delta_k \Omega_{k-1} \right)^{-1} \nonumber \\
&C^{(s+1)}=\left(\sum_{k=1}^{T} \Delta_k y_k \hat{x}_{k|T}^T\right)\left(\sum_{k=1}^{T} \Delta_k \Omega_k\right)^{-1}, \ \ P^{(s+1)} = \Omega_1 - \hat{x}_{1|T} \hat{x}_{1|T}^T \nonumber \\
&\Sigma^{(s+1)}=\frac{1}{T} \sum_{k=1}^{T}\Delta_k \left(y_k y_k^T - C^{(s+1)} \hat{x}_{k|T} y_k^T - y_k \hat{x}_{k|T}^T (C^{(s+1)})^T + C^{(s+1)} \Omega_k (C^{(s+1)})^T\right) \nonumber \\
&\Gamma^{(s+1)} = \frac{1}{T-1} \sum_{k=2}^{T} \frac{1}{\Delta_k}\left(\Omega_k-\Omega_{k,k-1}(A_k^{(s+1)})^T-A_k^{(s+1)}\Omega_{k,k-1}^T+A_k^{(s+1)}\Omega_{k-1}(A_k^{(s+1)})^{T}\right)
\end{align}
where $A_k^{(s)}=\text{Id}+\Delta_k A^{(s)}$.

\section{Clustering with mixtures of LGSSMs}   
\subsection{Mixture model}
We now move to extending the LGSSM of the previous section to the problem of identifying clusters within a dataset of multiple time series. Let us therefore assume that a population of $N$ time series $Y^i=(y_k^i)_{k=1,\ldots,T_i}$ are observed where for $i\in\{1,\ldots,N\}$, $T_i$ denotes the number of samples for time series $i$ and $0=t_1^i<\ldots<t_{T_i}^i$ are the corresponding time stamps. We shall extend the previous notation by defining $\Delta_k^i = t_{k}^i - t_{k-1}^i$ to be the sequence of time steps for the $i$-th time series.   

The underlying assumption leading to the proposed model is that the observed times series originate from independent observations of a mixture of $M$ LGSSMs, where $M$ is the presumed number of clusters to be identified. For $l=1,\ldots,M$, we will denote by $\theta_l = \{\mu_l,A_l,C_l,P_l,\Sigma_l,\Gamma_l\}$ the unknown parameters of the $l$-th cluster. To model the cluster membership of the different time series, we introduce a second set of latent variables $Z=\{z^1,\ldots,z^N\}$ where $z^i \in\{1,\ldots,M\}$ is a random variable that represents the index of the cluster assigned to subject $i$. The $z^i$'s are assumed to be independent and identically distributed with the unknown mixture probabilities $p(z^i = l) = \pi_l$ with $\sum_{l=1}^{M} \pi_l =1$. Thus the full set of parameters for the mixture model is $\Theta=\{(\theta_1,\pi_1),\ldots,(\theta_M,\pi_M)\}$.

For each $i\in\{1,\ldots,N\}$, we have $p(Y^i,X^i|z^i=l,\Theta) = p(Y^i,X^i|\theta_l)$ where the latter is given by the LGSSM of the previous section, i.e. by \eqref{eq:CDLL_LGSSM}. We can thus form the following complete data log-likelihood for the mixture model:
\begin{align*}
    \log p(X,Y,Z|\Theta) &= \sum_{i=1}^{N} \log p(X^i,Y^i,z^i|\Theta) \\
    &=\sum_{i=1}^{N} \log \prod_{l=1}^{M} \mathds{1}_{z^i=l} \, p(Y^i,X^i|\theta_l) \pi_l \\
    &=\sum_{i=1}^{N} \log p(Y^i,X^i|\theta_{z^i}) \pi_{z^i} =\sum_{i=1}^{N} \log p(Y^i,X^i|\theta_{z^i}) + \sum_{i=1}^{N} \log \pi_{z^i}
\end{align*}

\subsection{Expectation-Maximization clustering algorithm}
\subsubsection{E-step}
From the above complete data log-likelihood, we can extend the expectation-maximization scheme for LGSSMs discussed previously to the mixture model as follows. First, we note that the posterior distribution of the latent variables $(X,Z)$ satisfies:
\begin{equation*}
    p(X,Z|Y,\Theta) = p(Z|Y,\Theta) \times p(X|Y,Z,\Theta).
\end{equation*}
Furthermore, we have, on the one hand, that: 
\begin{equation*}
    p(X|Y,Z,\Theta)=\prod_{i=1}^N p(X^i|Y^i,z^i,\Theta) = \prod_{i=1}^N p(X^i|Y^i,\theta_{z^i})
\end{equation*}
and $p(X^i|Y^i,\theta_{z^i})$ is the posterior probability of the single LGSSM derived in the previous section. On the other hand, $p(Z|Y,\Theta) =\prod_{i=1}^{N} p(z^i|Y^i,\Theta)$.

Now, given the current estimate $\Theta^{(s)}$ of the parameters at iteration $s$, in the E-step, we can write the expected complete data log-likelihood function as:
\begin{equation}
\label{eq:ECDLL_MLGSSM1}
    Q(\Theta|\Theta^{(s)}) =  E_{p(Z|Y,\Theta^{(s)})} \left[E_{p(X|Y,Z,\Theta^{(s)})} \log p(X,Y,Z|\Theta) \right]
\end{equation}
Based on the above expressions, we first see that:
\begin{align*}
   E_{p(X|Y,Z,\Theta^{(s)})} \log p(X,Y,Z|\Theta) &= \sum_{i=1}^{N} E_{p\left(X^i|Y^i,\theta^{(s)}_{z^i}\right)} \left[\log p(Y^i,X^i|\theta_{z^i})\right] + \log \pi_{z^i} \\
   &=\sum_{i=1}^{N} Q_i(\theta_{z^i}|\theta_{z^i}^{(s)}) + \log \pi_{z^i}
\end{align*}
where $Q_i$ here denotes the expected complete data log-likelihood for the single LGSSM of the i-th observation. It then follows that \eqref{eq:ECDLL_MLGSSM1} becomes:
\begin{align}
\label{eq:ECDLL_MLGSSM2}
   Q(\Theta|\Theta^{(s)}) &= \sum_{i=1}^N \sum_{z^i=1}^{M} p(z^i|Y^i,\Theta^{(s)})(Q_i(\theta_{z^i}|\theta_{z^i}^{(s)}) + \log \pi_{z^i}) \nonumber\\
   &=\sum_{i=1}^N \sum_{l=1}^{M} p(z^i=l|Y^i,\Theta^{(s)})(Q_i(\theta_{l}|\theta_{l}^{(s)}) + \log \pi_{l})
\end{align}
and the posterior probabilities $p(z^i=l|Y^i,\Theta^{(s)})$, that we shall denote by $\tilde{\pi}^{(s)}_{l|Y^i}$, can be computed via Bayes rule and the law of total probability as follows:
\begin{align*}
    \tilde{\pi}^{(s)}_{l|Y^i}&=\frac{p(Y^i|z^i=l,\Theta^{(s)}) \, p(z^i=l)}{p(Y^i|\Theta^{(s)})} = \frac{p(Y^i|\theta_l^{(s)})\pi^{(s)}_l}{\sum_{u=1}^{M}p(Y^i|z^i=u,\Theta^{(s)})p(z^i=u) }\\&=\frac{p(Y^i|\theta_l^{(s)})\pi^{(s)}_l}{\sum_{u=1}^{M}p(Y^i|\theta_u^{(s)})\pi^{(s)}_u}.
\end{align*}
For each $u=1,\ldots,M$, $p(Y^i|\theta_u^{(s)})$ is given by evaluating \eqref{eq:prob_Y_theta} at $Y^i$. 

\subsubsection{M-step}
Let us now examine the maximization of \eqref{eq:ECDLL_MLGSSM2} with respect to the set of parameters $\Theta$. First, we focus on the mixture probabilities $\pi_l$. Given the constraint $\sum_{l=1}^{M} \pi_l = 1$, we consider the Lagrangian $\mathcal{L}(\Theta,\lambda) = Q(\Theta|\Theta^{(s)}) +\lambda\left(\sum_{l=1}^{M} \pi_l -1 \right)$. Then the maximum with respect to $\pi_m$ for $m=1,\ldots M$ must satisfy:
\begin{align*}
0&=\frac{\partial \mathcal{L}}{\partial \pi_m} = \frac{\partial}{\partial \pi_m}\sum_{i=1}^N \sum_{l=1}^{M} \tilde{\pi}^{(s)}_{l|Y^i}(Q_i(\theta_{l}|\theta_{l}^{(s)}) + \log \pi_{l}) + \lambda  \\
&=\frac{1}{\pi_m}\sum_{i=1}^N \tilde{\pi}^{(s)}_{m|Y^i} +\lambda 
\end{align*}
which implies that $\pi_m \lambda = -\sum_{i=1}^N \tilde{\pi}^{(s)}_{m|Y^i}$ and summing over $m$ yields:
\begin{equation*}
    \lambda = -\sum_{i=1}^N \underbrace{\sum_{l=1}^{M} \tilde{\pi}^{(s)}_{l|Y^i}}_{=1} = -N.
\end{equation*}
As a consequence, we find the following update for $\pi_l$:
\begin{equation}
\label{eq:M_step_MLGSSM_pi}
    \pi_l^{(s+1)} = \frac{1}{N} \sum_{i=1}^N \tilde{\pi}^{(s)}_{l|Y^i}
\end{equation}
As for the other parameters in $\Theta$, we have that for all $l=1,\ldots,M$:
\begin{equation*}
    0=\frac{\partial Q(\Theta|\Theta^{(s)})}{\partial \theta_l} = \sum_{i=1}^N \tilde{\pi}^{(s)}_{l|Y^i} \frac{\partial Q_i(\theta_{l}|\theta_{l}^{(s)})}{\partial \theta_l} 
\end{equation*}
Since $Q_i(\theta_{l}|\theta_{l}^{(s)})$ is the ECDLL of a single LGSSM, we can rely on the previous derivations for single LGSSMs. Using similar notations, we define $\hat{x}^{i,l}_{k|j}$, $\Omega_k^{i,l}$, $\Omega_{k,k-1}^{i,l}$ the a posteriori estimates of the single LGSSM for observation $i$ and parameters $\theta_l^{(s)}$.  We then get the following update equations:
\begin{align}
\label{eq:M_step_MLGSSM_theta}
    &\mu_l^{(s+1)} = \frac{\sum_{i=1}^{N} \tilde{\pi}^{(s)}_{l|Y^i} \hat{x}^{i,l}_{1|T}}{\sum_{i=1}^{N} \tilde{\pi}^{(s)}_{l|Y^i}} \nonumber\\
    &P_l^{(s+1)} = \frac{\sum_{i=1}^{N} \tilde{\pi}^{(s)}_{l|Y^i}\left[\Omega_1^{i,l} - \hat{x}^{i,l}_{k|T}(\mu_l^{(s+1)})^T - \mu_l^{(s+1)}(\hat{x}^{i,l}_{k|T})^T + \mu_l^{(s+1)} (\mu_l^{(s+1)})^T \right]}{\sum_{i=1}^{N} \tilde{\pi}^{(s)}_{l|Y^i}} \nonumber\\
    &A_l^{(s+1)} =\left(\sum_{i=1}^{N} \sum_{k=2}^{T_i} \tilde{\pi}^{(s)}_{l|Y^i}(\Omega_{k,k-1}^{i,l} - \Omega_{k-1}^{i,l}) \right) \left(\sum_{i=1}^{N} \sum_{k=2}^{T_i} \tilde{\pi}^{(s)}_{l|Y^i} \Delta_k^i \Omega_{k-1}^{i,l} \right)^{-1} \nonumber\\
    &C_l^{(s+1)}=\left(\sum_{i=1}^{N} \sum_{k=1}^{T_i} \tilde{\pi}^{(s)}_{l|Y^i} \Delta_k^i y_k^i (\hat{x}^{i,l}_{k|T})^T \right) \left(\sum_{i=1}^{N} \sum_{k=1}^{T_i} \tilde{\pi}^{(s)}_{l|Y^i} \Delta_k^i \Omega^{i,l}_k \right)^{-1} \nonumber\\
    &\Sigma_l^{(s+1)}=\sum_{i=1}^{N} \sum_{k=1}^{T_i} \tilde{\pi}^{(s)}_{l|Y^i} \Delta_k^i \big[y_k^i (y_k^i)^T -C_l^{(s+1)}\hat{x}^{i,l}_{k|T}(y_k^i)^T \nonumber\\ 
    &\hskip10ex -y_k^i(\hat{x}^{i,l}_{k|T})^T (C_l^{(s+1)})^T + C_l^{(s+1)} \Omega^{i,l}_k (C_l^{(s+1)})^T \big] \big / \left(\sum_{i=1}^{N} \tilde{\pi}^{(s)}_{l|Y^i} T_i\right) \nonumber\\
    &\Gamma_l^{(s+1)}= \sum_{i=1}^{N} \sum_{k=2}^{T_i} \frac{\tilde{\pi}^{(s)}_{l|Y^i}}{\Delta_k^i} \big[\Omega_k^{i,l} - \Omega_{k,k-1}^{i,l}(A_{l,k}^{(s+1)})^T - A_{l,k}^{(s+1)} (\Omega_{k,k-1}^{i,l})^T \nonumber\\
    &\hskip15ex + A_{l,k}^{(s+1)} \Omega_{k,k-1}^{i,l} (A_{l,k}^{(s+1)})^T\big] \big/ \left(\sum_{i=1}^{N} \tilde{\pi}^{(s)}_{l|Y^i} (T_i-1) \right)
\end{align}

\subsubsection{Cluster assignment}
The EM algorithm is terminated once the difference between two consecutive sets of parameters $\Theta^{(s)}$ and $\Theta^{(s+1)}$ falls below a given threshold. The cluster index $z_i^* \in \{1,\ldots,M\}$ assigned to each observed time series $Y^i$ is then determined based on the final posterior probabilities $\tilde{\pi}^{(s)}_{l|Y^i}$ according to the following rule:
\begin{equation}
    \label{eq:cluster_assignment}
    z_i^* = \underset{l=1,\ldots,M}{\operatorname{argmax}} \ \tilde{\pi}^{(s)}_{l|Y^i}
\end{equation}
Note that although we choose to select deterministic cluster assignments for the results that are shown in the next section, the estimated model further provides uncertainty measures on those assignments based on the computed probabilities $\{\tilde{\pi}^{(s)}_{l|Y^i}\}$.

\subsection{Implementation details}

\subsubsection{Algorithm}\label{ssec:algorithm}
 The previous mixture LGSSM clustering approach was implemented in Python, building from the EM implementation introduced in \parencite{umatani2023time}. VISTA takes as input the data $Y$, the time stamps $t^i_k$ for each observation in each time series, the presumed number of clusters $M$, the predetermined dimension $d$ of the latent variable, a termination tolerance $\varepsilon$, the maximum number of iterations to run the algorithm (should the tolerance $\varepsilon$ not be met), the number of CPUs to use, and the method of initialization with associated hyperparameters. The initialization strategies are discussed in the following section. Since many of the parameter updates given by \eqref{eq:M_step_MLGSSM_theta} involve double sums over the $N$ time series with varying number $T_i$ of time samples, we sought to minimize the computational burden of these calculations. The implementation developed in \parencite{umatani2023time} assumes an identical number of time samples across the population and precomputes each $\hat{x}^{i,l}_{k|j}$, $\Omega_k^{i,l}$, and $\Omega_{k,k-1}^{i,l}$ for all samples ($k$) in parallel over all time series ($i$) and clusters ($l$) to generate arrays of sizes $N \cross M \cross T \cross d$, $N \cross M \cross T \cross d \cross d$, and $N \cross M \cross T-1 \cross d \cross d$ respectively, in order to compute \eqref{eq:M_step_MLGSSM_theta} with linear algebra array operations. In contrast, we forego unwieldy large arrays, which pose the risk of causing memory issues for large datasets, and instead tackle the sums in \eqref{eq:M_step_MLGSSM_theta} online in parallel over all time series. During the summing process we run the Kalman filter and RTS smoother to compute $\hat{x}^{i,l}_{k|j}$, $\Omega_k^{i,l}$, and $\Omega_{k,k-1}^{i,l}$ for each time series $i$ in much the same manner as \parencite{umatani2023time}, with modifications to include the parameters being a function of the timestep due to our discretization. Prior to the summing process we compute $\tilde{\pi}^{(s)}_{l|Y^i}$ for all time series and clusters, saving the information in a $N \cross M$ array; this is achieved by calculating $\log(p(Y^i|\theta_u^{(s)})\pi^{(s)}_u)$ for each cluster $u$ and time series $i$ then later transforming it with the exponential function to compute $\tilde{\pi}^{(s)}_{l|Y^i}$ without risk of overflow. With these steps and the previous parameter values we perform the M-step updates outlined above at each iteration $s$ of the EM algorithm. The algorithm is terminated when the sum over the absolute difference between $\{\theta_1^{(s+1)},\ldots,\theta_M^{(s+1)}\}$ and $\{\theta_1^{(s)},\ldots,\theta_M^{(s)}\}$ (summing over each entry in each matrix parameter) is less than a chosen tolerance $\varepsilon$. In our experiments we take $\varepsilon = 0.1$. This is summarized at a high level in both figure \ref{fig:schematic}, which presents a schematic of the algorithm, and algorithm \ref{alg:vista}.

 \begin{figure}[t]
    \begin{algorithm}[H]
        \caption{VISTA}
        \label{alg:vista}
        \begin{algorithmic}[1]
            \REQUIRE Data $Y$ of $N$ time series, number of clusters $M$, dimension of latent space $d$, initialization method, termination tolerance $\varepsilon_0$, and maximum number of iterations $maxiter$
            \STATE Compute initial parameters $\Theta^{(0)}$ from inputs
            \FOR{$s=0,\;\dotsc\,,\;maxiter$}
            \STATE // E-step
            \FOR{$i=1,\;\dotsc\,,\;N$} 
            \FOR{$k=1,\;\dotsc\,,\;M$}
            \STATE Compute $\hat x^{i,l}_{k|j}$, $\Omega^{i,l}_k$, and $\Omega^{i,l}_{k,k-1}$ with the Kalman filter and RTS smoother \eqref{eq:Kalman_filtering1}, \eqref{eq:Kalman_filtering2}, \eqref{eq:Kalman_smoothing}
            \STATE Compute the expected complete data log-likelihood function $Q(\Theta | \Theta^{(s)})$ \eqref{eq:ECDLL_MLGSSM1}, \eqref{eq:ECDLL_MLGSSM2}
            \ENDFOR
            \FOR{$k=1,\;\dotsc\,,\;M$}
            \STATE Compute posterior probabilities $\tilde{\pi}^{(s)}_{l|Y^i}$
            \ENDFOR
            \ENDFOR
            \STATE // M-Step
            \FOR{$k=1,\;\dotsc\,,\;M$}
            \STATE Compute \eqref{eq:M_step_MLGSSM_pi}, \eqref{eq:M_step_MLGSSM_theta} to determine $\Theta^{(s+1)}$
            \ENDFOR
            \STATE // Convergence criterion
            \STATE $\varepsilon \leftarrow \bigl\|\Theta^{(s+1)} - \Theta^{(s)} \bigr\|_1$
            \IF {$\varepsilon < \varepsilon_0$}
            \RETURN $\Theta^{(s+1)}$
            \ENDIF
            \ENDFOR
            \RETURN labels $z^*_i$ \eqref{eq:cluster_assignment} for each $i=1,\;\dotsc\,,\;N$ and parameters $\Theta^{(s+1)}$
        \end{algorithmic}
    \end{algorithm}
\end{figure}

 Lastly, we note that VISTA allows for some of the parameters of the mixture model to be fixed while the others are optimized by the algorithm. It is simply done by restricting the above updates to a given subset of the parameter list in $\Theta$. This can be especially relevant in  situations where reasonable prior estimates for certain parameters can be used and the amount of available data may be too low with respect to the dimension of the full set of parameters.

\subsubsection{Initialization}\label{ssec:initialization}

Although EM schemes are designed so that the expected posterior expectation $Q(\Theta|\Theta^{(s)})$ is non-decreasing throughout iterations, they are not guaranteed to converge to a true global maximum of the actual log-likelihood. Thus the choice of initialization of the parameters usually plays a critical role in getting the EM to converge to an acceptable local maximum of the likelihood. We provide and experiment with three different methods for initializing VISTA. The first method, which we refer to as the \textit{identity initialization}, consists in the following fixed set of parameters for each cluster $l$:
\begin{equation}\label{eq:ident_init}
\begin{split}
    \mu_l^{(0)} = (-1+2\frac{l-1}{M-1}) \cross \mathbf{1}, \quad
    P_l^{(0)} = 0.1 \cross \text{Id}, \quad
    A_l^{(0)} = -1.5 \cross \text{Id}, \quad \\
    \Gamma_l^{(0)} = 0.1 \cross \text{Id}, \quad
    C_l^{(0)} = \mathbf{C}, \quad
    \Sigma_l^{(0)} = 0.1 \cross \text{Id}, \quad
    \pi_l^{(0)} = 1/M
\end{split}
\end{equation}
where $\mathbf{1}$ is the array of all ones of the appropriate dimension for its corresponding assignment and $\mathbf{C}$, in the case of $d \neq n$, is comprised of an identity matrix in the smaller dimension, with the remaining entries comprised of rows or columns of the same dimension identity matrix to fill out the matrix.

A second possible approach is \textit{random initialization}, namely one sets for cluster $l$:
\begin{equation}\label{eq:ident_random}
\begin{split} 
    \mu_l^{(0)} = \bar \mu_l, \quad
    P_l^{(0)} \in \text{PSD}_d, \quad
    A_l^{(0)} = \bar A_l, \quad \\
    \Gamma_l^{(0)} \in \text{PSD}_d, \quad
    C_l^{(0)} = \bar C_l, \quad
    \Sigma_l^{(0)} \in \text{PSD}_n, \quad
    \pi_l^{(0)} = 1/M
\end{split}
\end{equation}
where for $l=1,\ldots,M$, $\bar \mu_l$ has entries chosen uniformly at random from $[0,1)$ and $\text{PSD}_n$ denotes the set of symmetric positive definite matrices of size $n \cross n$. Each $\bar C_l$ is constructed by setting all elements in the first row to 1 and choosing the other entries to be uniformly either 0 or 1, while each $\bar A_l$ is diagonal with entries chosen uniformly at random from $[-1.9,-0.1)$, which guarantees each $A_k$ will have eigenvalues bounded in modulus by 1 to ensure asymptotic stability for the LGSSM \eqref{eq:lgssm}.

The final method is a modification of the \textit{k-means initialization} proposed in \parencite{umatani2023time}. In this approach, one begins by fitting individual LGSSMs to each of the time series in the dataset via the EM algorithm for a single LGSSM given previously with a common initialization given by:
\begin{equation}\label{eq:kmeans_init}
\begin{split}
            \mu^{(0)} = \mathbf{0}, \quad
    P^{(0)} = 10^4 \cross \text{Id}, \quad
    A_k^{(0)} = O, \quad \\
    \Gamma_k^{(0)} = 0.05 \cross \text{Id}, \quad
    C^{(0)} = \mathbf{1}, \quad
    \Sigma_k^{(0)} = 0.05 \cross \text{Id}
\end{split}
\end{equation}
where $\mathbf{0}$ is the array of all zeros of the appropriate dimension for its corresponding assignment and $O$ an orthogonal matrix obtained from QR decomposition of a standard Gaussian random matrix. For each time series, a predetermined number of initializations (chosen as 30 in our experiments) are generated and LGSSMs are fit from each initialization. As observed by \parencite{steinley2006k, steinley2007initializing}, the k-means algorithm can take thousands of different random initializations to reach a globally optimal partition due to large numbers of locally optimal partitions. As we use k-means only to determine an initialization for VISTA, we are less concerned with finding a globally optimal partition, as VISTA itself will develop the model parameters further than the k-means initialization. The only parameter varying among each of these initializations within the k-means initialization method is $A_k^{(0)}$, following the initialization method in \parencite{umatani2023time} with the intent to describe as much of the signal as possible through the evolution of the latent variable, rather than attributing it to noise. The LGSSM that generates the largest log-likelihood (computed as $\log p(Y^i|\theta^i)$ from \eqref{eq:prob_Y_theta}, where $\theta^i$ represents the terminal parameter values given by the EM algorithm for a single LGSSM for the time series $i$) is chosen for that time series, yielding a set of parameters $\hat \theta^i$ for each time series $i$. Ultimately, we vectorize each $\hat \theta^i$, such that each parameter set is a single vector. We then perform k-means clustering over all $\hat \theta^i$, returning both the centers and proportions of class membership, which can be converted to the parameter set $\Theta^{(0)}$ used for initialization by reversing the vectorization process.

As a general rule, the latter initialization strategy is usually preferred in practice as it empirically leads to higher final log-likelihood values. However, it is typically less well suited in situations where the number of parameters in a single LGSSM is too large compared to the number of time observations in each time series.

\section{Evaluation Studies}

We evaluated VISTA on a variety of examples, including simulated data and publicly available datasets consisting of common use cases in psychology and healthcare. These include population survey trends, wearable data signals from an epilepsy dataset, epidemiological time series from COVID-19, and ecological momentary assessments of emotional states and depressive symptoms. In each case we rescaled the timestamps to [0,1] to establish consistency across our experiments. Each study is described in detail in their respective sections, and the datasets and code have been made publicly available in our repository \footnote{\url{https://github.com/benjaminbrindle/vista_ssm}}.

We benchmarked VISTA against different state-of-the-art time series analysis approaches. These include feature-based methods such as the dynamic time warping k-means (DTW) and residual neural net (ResNet) clustering methods that are part of the Aeon library\footnote{\url{https://github.com/aeon-toolkit/aeon/}}, clustering via discrete wavelet transform \footnote{\url{https://www.mathworks.com/help/wavelet/ref/mdwtcluster.html}} as well as model-based clustering based on the vector autoregressive framework of \parencite{bulteel2016clustering}.   

Unlike supervised tasks such as classification, measuring the quality of clustering results for benchmarking can be challenging in the absence of actual ground truth classes. For this reason, we focused our experiments on datasets with available ground truths that divide the population into relatively distinct groups, with the exception of the COVID-19 example. Clustering accuracy can then be evaluated by measuring the fraction of overlap between the extracted clusters with the ground truth groups. However, the labels returned by an unsupervised algorithm may not correspond directly to the ground truth labels as the labels could be permuted even when perfectly corresponding. We overcome this obstacle by creating a confusion matrix between the true labels and the cluster labels and determining the optimal permutation of cluster labels with respect to maximizing the trace of the confusion matrix, causing the most overlap between the true classes and the clusters. For each of our experiments with ground truth labels, we provide the clustering accuracy defined above as well as the Adjusted Rand Index \parencite{hubert1985comparing} to evaluate the performance of VISTA. In situations where no ground truth labels exist, the meta parameters $M$ and $d$ with which to run VISTA must be selected. For model selection, we use the adjusted Bayesian Information Criterion (ABIC) \parencite{sclove1987application} based on the findings of \parencite{nylund2007deciding} that the ABIC more consistently recovers the true number of classes than other information criteria in the latent class analysis setting. The ABIC is defined as
\[\text{ABIC} =  -2 \log L + p \log \qty(\frac{N+2}{24})\]
where $L$ represents the maximized likelihood of the model and $p$ the number of free parameters. Models with lower ABIC values are preferred.

\subsection{Simulated Data}

We simulated a dataset of time series generated from discrete dynamical systems of the form \eqref{eq:lgssm}, using three separate groups of parameters to create a three class problem. The dataset takes $d=5$ and $n=2$ for 2 observed dimensions. The classes are created unbalanced, comprised of 40, 50, and 30 time series, respectively. Each time series has length chosen uniformly at random from 25 to 75, inclusively. The time stamps are generated for the $i$th time series by setting $t^i_1=0$ then choosing the increments $\Delta^i_k$, $k \in \{2,..,T_i\}$, uniformly at random from 1 to 5, inclusively. Finally, the time stamps are divided by $t^i_{T_i}$, the terminal time, to rescale the time stamps to lie within $[0,1]$ for every time series.

\begin{figure}[H]
    \centering
    \includegraphics[width=\linewidth]{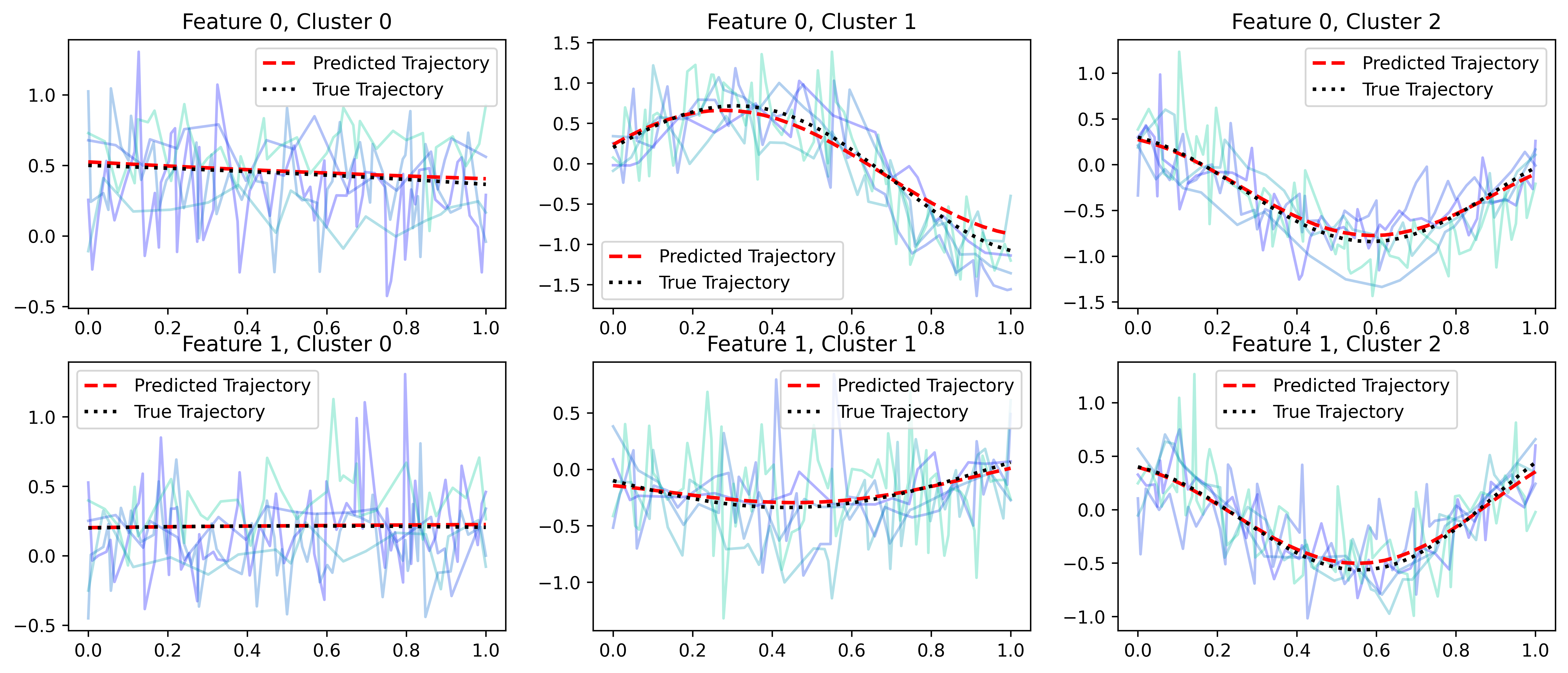}
    \caption{The result of VISTA with the returned clusters and noiseless trajectory determined by the optimal parameters $\mu$, $A$, and $C$ compared to the noiseless trajectory determined from the parameters used to generate the data along with five arbitrarily chosen time series from the simulated dataset with $d=5$ and $n=2$.}
    \label{fig:sim1}
\end{figure}

We run VISTA on this simulated dataset for a variety of values of $M$ and $d$ and analyze the ABIC to determine the best model. ABIC values returned from running VISTA with the identity initialization on each $M$ and $d$ pair, chosen for reproducability, are shown in table \ref{tab:sim}. We observe the ABIC is lowest when $M=3$ and $d=5$, recovering the underlying parameters from the mixture of LGSSMs used to generate the data. With $M=3$ and $d=5$ we achieve perfect cluster similarity, as exhibited in the confusion matrix in table \ref{cm:sim1}. Despite the considerable noise in the data shown in figure \ref{fig:sim1}, VISTA is able to recover the ground truth class labels and produce a predicted trajectory that closely matches that of the underlying dynamical system used to generate the data.

\begin{table}[H]
\centering
\caption{ABIC from identity initialization for varying $M$ and $d$.}
\begin{tabular}{|c| c| c|c|}
\hline
\diagbox{$d_x$}{$M$} & 2 & 3 & 4 \\ \hline
3 & 3.83E+03 & 3.11E+03 & 2.90E+03  \\ \hline
4 & 4.09E+03 & 3.27E+03 & 3.48E+03  \\ \hline
5 & 3.95E+03 & \textbf{2.75E+03} & 2.88E+03  \\ \hline
6 & 4.07E+03 & 3.52E+03 & 3.10E+03  \\ \hline
7 & 4.22E+03 & 3.98E+03 & 3.36E+03  \\ \hline
\end{tabular}
\label{tab:sim}
\end{table}

\begin{table}[H]
\centering
\caption{Confusion matrix for the simulated dataset with $d=5$ and $n=2$.}
\begin{tabular}{|l|l|l|l|}
\hline
\diagbox{True}{Predicted} & Cluster 0 & Cluster 1 & Cluster 2 \\ \hline
Cluster 0  & 40        & 0         & 0         \\ \hline
Cluster 1  & 0         & 50        & 0         \\ \hline
Cluster 2 & 0         & 0         & 30        \\ \hline
\end{tabular}
\label{cm:sim1}
\end{table}

\subsection{Panel Data: State Population} 
\begin{figure}[H]
    \centering
    
        \begin{subfigure}[c]{\linewidth}
            \centering
            \includegraphics[width=\linewidth]{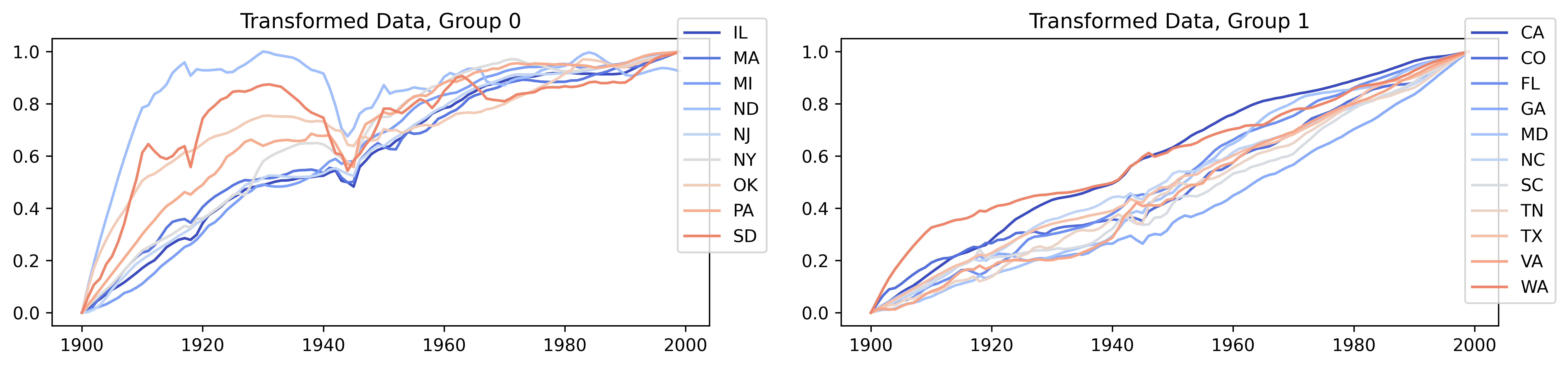}
            \subcaption{Log-transformed and min-maxed state population time series. For each state, 0 corresponds to the minimum log population value, and 1 the maximum log population value. The raw data in group 0 exhibited a linear trend, while group 1 exhibited an exponential trend - corresponding to the logarithmic and linear trends here.}
            \label{pop1}
        \end{subfigure}
                \begin{subfigure}[c]{\linewidth}
            \centering
              \includegraphics[width=\linewidth]{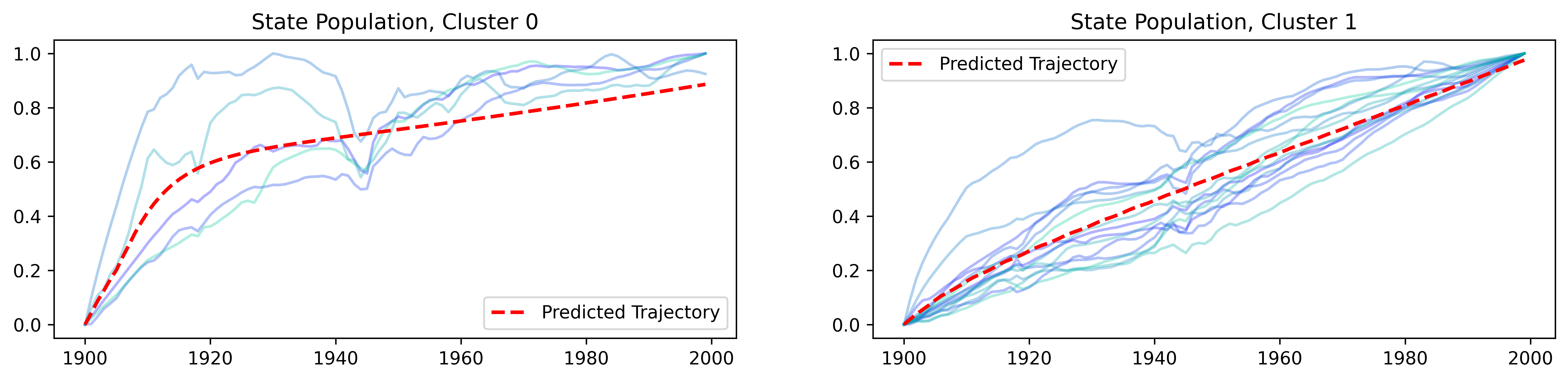}
            \subcaption{Clusters returned by VISTA using the k-means initialization. Predicted trajectories are plotted by using the optimal parameters returned from VISTA on the LGSSM equations while setting the covariance matrices equal to zero to remove noise.}
            \label{pop2}
        \end{subfigure} 

    \caption{Population time series. (a) shows the observation data fed to VISTA, split into ground truth groups. (b) shows the result of VISTA with the returned clustered and noiseless trajectory determined by the optimal parameters $\mu$, $A$, and $C$.}
    \label{populationplots}
\end{figure}

The state population dataset\footnote{\url{https://web.archive.org/web/20040220002039/https://eire.census.gov/popest/archives/state/st_stts.php}, also available at \url{https://github.com/benjaminbrindle/vista_ssm} for easy access.} examines the evolution of the populations of twenty states in the United States yearly from 1900 to 1999. Class labels are created by dividing the states into groups depending on whether their population growth was approximately linear (9 states) or approximately exponential (11 states). To remove the influence of vastly differing population sizes among the states (contrast North Dakota's 644 thousand with New York's 18,833 thousand in 1999), we normalize the time series by first log-transforming the raw data, then min-max normalize each time series such that for each state the minimum value becomes 0 and the maximum value becomes 1, as can be seen in figure \ref{pop1}. 

We demonstrate VISTA's performance on this classification problem using each of the initialization methods discussed in the initialization section above. First, we show analysis of the ABIC in table \ref{population:bic}. To determine these scores, we run the algorithm with the k-means initialization 10 times for each value of $d$ shown and display the average ABIC. Likewise, we run VISTA 10 times for the random initialization. As the identity initialization is deterministic, we only need to run it once. 

In table \ref{population:iter}, we display the average number of iterations taken by VISTA to terminate by either converging or reaching 1000 iterations for each initialization method. We observe that the identity initialization converges in the fewest number of iterations, but see in table \ref{population_tables} that this comes at the expense of a high ABIC and poor performance in cluster similarity and ARI. The random initialization, on the other hand, takes by far the most iterations to terminate while producing the lowest ABIC and best cluster similarities and ARI. However, we notice that the k-means initialization generates comparable results in ABIC, cluster similarity, and ARI in far fewer iterations. In the setting of the panel data, the computational expense of the large number of iterations required for the random initialization is offset by the small size of the dataset. However, for larger datasets like the wearable data examined below, running VISTA for a large number of iterations quickly becomes impractical, making the k-means initialization attractive. Alternatively, for situations like the simulated data where the number of parameters for each LGSSM is large compared to the number of observations in an individual time series, the k-means initialization may prove ill-posed. With these observations in mind, we elect to examine the k-means initialization for our remaining analysis of the panel data to remain consistent in our overall methodology. 

We observe that, for the k-means initialization, the lowest ABIC score occurs when $d=12$. By referencing the ground truth labels, we display the cluster similarity in table \ref{population:accuracy} and ARI in table \ref{population:ari} - the choice of $d=12$ and the k-means initialization results in one of the better average cluster similarities at 0.82. 

\begin{table}[H]
    \centering
    \caption{
        An example confusion matrix for $d=12$ and the k-means initialization corresponding to the best average ABIC score for the panel data along with the mean (standard deviation) number of iterations taken before termination for each initialization method.
    }
    \label{population_tables2}
    \begin{tabular}{cc}

        \begin{minipage}{0.6\linewidth}
            \centering
            \subcaption{Typical Confusion Matrix, $d=12$, k-means initialization}
            \label{population:cm}
            \begin{tabular}{|c|c|c|}
                \hline
                \diagbox{True}{Predicted} & \makecell{Cluster 0\\(Logarithmic)} & \makecell{Cluster 1\\(Linear)} \\ 
                \hline
                \makecell{Group 0\\(Logarithmic)} & 7 & 2 \\ 
                \hline
                \makecell{Group 1\\(Linear)} & 0 & 11 \\ 
                \hline
            \end{tabular}
        \end{minipage}

        &
        
        \begin{minipage}{0.33\linewidth}
            \centering
            \subcaption{Mean number of iterations taken before termination}
            \label{population:iter}
            \begin{tabular}{|c|c|}
                \hline
                Method & Iterations \\ 
                \hline
                k-means & \makecell{59.35\\(90.09)} \\ 
                \hline
                random & \makecell{254.35\\(233.17)} \\ 
                \hline
                identity & \makecell{9.15\\(1.17)} \\ 
                \hline
            \end{tabular}
        \end{minipage}
        
    \end{tabular}
\end{table}

In table \ref{population:cm} we display the best confusion matrix for $d=12$ and the k-means initialization; this particular confusion matrix corresponds to a cluster similarity of 0.9. In figure \ref{pop2}, we display the time series placed in each cluster by VISTA with the k-means initialization when $d=12$, along with a noiseless trajectory predicted from the LGSSM parameters returned by VISTA. We observe a strong logarithmic trend of both the time series and predicted trajectory in cluster 0. In cluster 1 , we notice a linear trend, with some of the more weakly logarithmic states having been misclassified here.

\begin{table}[H]
    \caption{
        Mean (standard deviation) of ABIC (in ten thousands), cluster similarity, and ARI from 10 trials with k-means and random initializations and 1 trial with identity initialization for $M=2$ clusters and fixed $d$ for the panel data.
    }
    \begin{tabular}{c}
        
        \begin{minipage}[c]{\linewidth}
            \centering
            \subcaption{Adjusted Bayesian Information Criterion (in ten thousands)}
            \label{population:bic}
             \resizebox{\textwidth}{!}{
\begin{tabular}{|p{0.10\linewidth} | p{0.07\linewidth} |p{0.07\linewidth} |p{0.07\linewidth} |p{0.07\linewidth} |p{0.07\linewidth} |p{0.07\linewidth} |p{0.07\linewidth} |p{0.07\linewidth} |p{0.07\linewidth} |p{0.07\linewidth} |p{0.07\linewidth} |p{0.07\linewidth} |p{0.07\linewidth}|}
\hline
$d$  & 2                  & 3                  & 4                  & 5                  & 6                  & 7                  & 8                  & 9                  & 10                 & 11                 & 12                 & 13                 & 14                 \\ \hline
k-means & -1.26 (0.027) & -1.25 (0.038) & -1.28 (0.010) & -1.28 (0.018) & -1.27 (0.017) & -1.28 (0.018) & -1.28 (0.014) & -1.29 (0.011) & -1.29 (0.013) & -1.28 (0.039) & -1.30 (0.014) & -1.29 (0.016) & -1.29 (0.019)
 \\ \hline
random & -1.27 (0.021) & 1.29 (0.003) & -1.30 (0.025) & -1.31 (0.003) & -1.31 (0.008) & -1.31 (0.004) & -1.31 (0.010) & -1.32 (0.003) & -1.32 (0.008) & -1.32 (0.012) & -1.33 (0.015) & -1.33 (0.015) & -1.33 (0.013)
\\ \hline
identity & -1.05 & -1.04 & -1.04 & -1.04 & -1.04 & -1.04 & -1.07 & -1.09 & -1.08 & -1.08 & -1.08 & -1.10 & -1.10
\\ \hline
\end{tabular} 
}
            \vspace{5mm}
        \end{minipage}

\\

        \begin{minipage}[c]{\linewidth}
            \centering
            \subcaption{Cluster Similarity}
            \label{population:accuracy}
                         \resizebox{\textwidth}{!}{
\begin{tabular}{|p{0.10\linewidth} | p{0.07\linewidth} |p{0.07\linewidth} |p{0.07\linewidth} |p{0.07\linewidth} |p{0.07\linewidth} |p{0.07\linewidth} |p{0.07\linewidth} |p{0.07\linewidth} |p{0.07\linewidth} |p{0.07\linewidth} |p{0.07\linewidth} |p{0.07\linewidth} |p{0.07\linewidth}|}
\hline
$d$        & 2           & 3           & 4           & 5           & 6           & 7           & 8           & 9           & 10          & 11          & 12          & 13          & 14          \\ \hline
k-means & 0.84 (0.06) & 0.78 (0.08) & 0.82 (0.05) & 0.81 (0.03) & 0.79 (0.05) & 0.80 (0.06) & 0.80 (0.04) & 0.78 (0.05) & 0.79 (0.05) & 0.80 (0.04) & 0.82 (0.05) & 0.80 (0.05) & 0.79 (0.07) \\ \hline 
random & 0.81 (0.05) & 0.82 (0.04) & 0.83 (0.05) & 0.85 (0.00) & 0.85 (0.00) & 0.85 (0.00) & 0.85 (0.00) & 0.85 (0.00) & 0.84 (0.01) & 0.85 (0.00) & 0.85 (0.00) & 0.85 (0.00) & 0.85 (0.00) \\ \hline
identity & 0.65 & 0.65 & 0.65 & 0.65 & 0.65 & 0.65 & 0.65 & 0.65 & 0.65 & 0.65 & 0.65 & 0.65 & 0.65 \\ \hline
\end{tabular}
}
            \vspace{5mm}
        \end{minipage}
\\
        \begin{minipage}[c]{\linewidth}
            \centering
            \subcaption{Adjusted Rand Index}
            \label{population:ari}
                         \resizebox{\textwidth}{!}{
\begin{tabular}{|p{0.10\linewidth} | p{0.07\linewidth} |p{0.07\linewidth} |p{0.07\linewidth} |p{0.07\linewidth} |p{0.07\linewidth} |p{0.07\linewidth} |p{0.07\linewidth} |p{0.07\linewidth} |p{0.07\linewidth} |p{0.07\linewidth} |p{0.07\linewidth} |p{0.07\linewidth} |p{0.07\linewidth}|}
\hline
$d$        & 2           & 3           & 4           & 5           & 6           & 7           & 8           & 9           & 10          & 11          & 12          & 13          & 14          \\ \hline
k-means & 0.47 (0.18) & 0.31 (0.19) & 0.39 (0.14) & 0.36 (0.08) & 0.33 (0.13) & 0.33 (0.13) & 0.33 (0.10) & 0.31 (0.10) & 0.32 (0.13) & 0.34 (0.11) & 0.38 (0.13) & 0.36 (0.12) & 0.34 (0.15)
 \\ \hline 
random & 0.37 (0.13) & 0.39 (0.12) & 0.42 (0.10) & 0.47 (0.00) & 0.47 (0.00) & 0.47 (0.00) & 0.47 (0.00) & 0.46 (0.00) & 0.45 (0.04) & 0.47 (0.00) & 0.47 (0.00) & 0.47 (0.00) & 0.47 (0.00) \\ \hline
identity & 0.07 &  0.07 &  0.07 &  0.07 &  0.07 &  0.07 &  0.07 &  0.07 &  0.07 &  0.07 &  0.07 &  0.07 &  0.07 \\ \hline
\end{tabular}
}
        \end{minipage}

    \end{tabular}
    \label{population_tables}
\end{table} 

The average cluster similarity of 0.82 returned from the information criterion analysis is consistent with \parencite{umatani2023time} and the maximum cluster similarity of 0.9 performs consistently with other classification methods shown in table \ref{comparisons}. However, we notice from table \ref{population:accuracy} that other choices of $d$ return higher average cluster similarities, showing the potential of our algorithm to accurately recover ground truth labels on noisy data.

\subsection{Wearable Data: Epilepsy Seizures} 

The epilepsy dataset\footnote{Collected from the aeon Python package \url{https://web.archive.org/web/20221012032341/http://timeseriesclassification.com/description.php?Dataset=Epilepsy}} \parencite{villar2016generalized} is comprised of 275 time series of 206 time points taken from a tri-axial accelerometer recording four distinct actions of six subjects, sampling at 16Hz - resulting in three features. The four actions correspond to a four class problem. The subjects are either walking, running, sawing, or imitating a seizure. The original data was preprocessed to ensure all time series would be of equal length by truncating data to the length of the shortest time series.

We demonstrate VISTA's performance on this clustering problem using the k-means initialization method. First, we show ABIC analysis in table \ref{epilepsy:bic}. To determine these scores, we run VISTA 10 times for each value of $d$ shown and display the average ABIC. In this four-class setting, the lowest ABIC occurs when $d=14$. We show the cluster similarity and ARI in table \ref{epilepsy:accuracy}. In table \ref{epilepsy:cm} we display the confusion matrix of the execution of VISTA that returned the highest cluster similarity for $d=14$; the cluster similarity in this case is 0.953.

\begin{table}[H]
    \caption{
        Mean (standard deviation) of ABIC (in ten thousands) and cluster similarity from 10 trials with k-means initialization for $M=4$ clusters and fixed $d$.
    }
    \begin{tabular}{c}
        
        \begin{minipage}[c]{\linewidth}
            \centering
            \subcaption{Adjusted Bayesian Information Criterion (in ten thousands)}
            \label{epilepsy:bic}
             \resizebox{\textwidth}{!}{
\begin{tabular}{|p{0.07\linewidth} | p{0.08\linewidth} |p{0.08\linewidth} |p{0.08\linewidth} |p{0.08\linewidth} |p{0.08\linewidth} |p{0.08\linewidth} |p{0.08\linewidth} |p{0.08\linewidth} |p{0.08\linewidth} |p{0.08\linewidth} |p{0.08\linewidth} |p{0.08\linewidth}|}

\hline
$d$  & 3        & 4        & 5        & 6        & 7        & 8        & 9        & 10       & 11       & 12       & 13       & 14       \\ \hline
ABIC & 19.4 (0.6849)     & 15.0 (1.1953)    & 11.3 (0.8155)     & 9.25 (0.6402)     & 8.17 (1.3159)    & 7.20 (1.3116)    & 7.35 (2.3141)    & 5.21 (0.9343)      & 5.38 (1.3587)     & 5.18 (1.4423)     & 5.72 (1.7633)     & 4.50 (1.9543) \\ \hline
\end{tabular}
}
            \vspace{5mm}
        \end{minipage}

        \\

        \begin{minipage}[c]{\linewidth}
            \centering
            \subcaption{Cluster Similarity and Adjusted Rand Index}
            \label{epilepsy:accuracy}
\resizebox{\textwidth}{!}{
\begin{tabular}{|p{0.10\linewidth} | p{0.07\linewidth} |p{0.07\linewidth} |p{0.07\linewidth} |p{0.07\linewidth} |p{0.07\linewidth} |p{0.07\linewidth} |p{0.07\linewidth} |p{0.07\linewidth} |p{0.07\linewidth} |p{0.07\linewidth} |p{0.07\linewidth} |p{0.07\linewidth}|}
\hline
$d$        & 3           & 4           & 5           & 6           & 7           & 8           & 9           & 10          & 11          & 12          & 13          & 14          \\ \hline
Similarity & 0.79 (0.10) & 0.76 (0.10) & 0.82 (0.08) & 0.87 (0.08) & 0.83 (0.10) & 0.84 (0.11) & 0.85 (0.11) & 0.90 (0.06) & 0.85 (0.09) & 0.84 (0.10) & 0.80 (0.10) & 0.86 (0.10) \\ \hline
ARI & 0.63 (0.12) & 0.59 (0.12) & 0.66 (0.08) & 0.74 (0.11) & 0.69 (0.13) & 0.69 (0.17) & 0.72 (0.12) & 0.78 (0.09) & 0.71 (0.11) & 0.70 (0.11) & 0.66 (0.11) & 0.72 (0.13) \\ \hline
\end{tabular}
}
        \end{minipage}

    \end{tabular}
    \label{epilepsy_tables}
\end{table}

\begin{table}[H]
            \centering
            \caption{Confusion matrix for $M=4$ and $d=14$ corresponding to the highest cluster similarity (0.953) from the best average ABIC.}
            \label{epilepsy:cm}
            \begin{tabular}{|c|c|c|c|c|} \hline
                \diagbox{True}{Predicted} & Epilepsy & Walking & Running & Sawing \\ \hline Epilepsy & 63&5&0&0\\ \hline
                Walking & 0&69&5&0\\ \hline
                Running & 0&0&73&0\\ \hline
                Sawing & 0&3&0&57 \\ \hline
            \end{tabular}
\end{table}

VISTA handles this four-class classification problem well, averaging 0.85 or better for many of the choices of $d$ shown in table \ref{epilepsy:accuracy}, while other methods for time series clustering shown in table \ref{comparisons} achieve scores of less than 0.50. Cochran's Q test \parencite{cochran1950comparison} confirms this performance gap is statistically significant (Q = 46.29, df = 4, p $<$ 0.0001).

\begin{table}[H]
\caption[caption]{Cluster similarities and ARIs obtained by several unsupervised time series clustering methods compared against VISTA. We use both k-means with dynamic time warping (DTW) and an auto-encoder with a residual neural network (ResNet) from the aeon toolkit\tablefootnote{\url{https://github.com/aeon-toolkit/aeon/}}, vector autoregressive clustering \parencite{bulteel2016clustering} (VAR), and discrete wavelet transform from MATLAB\tablefootnote{\url{https://www.mathworks.com/help/wavelet/ref/mdwtcluster.html}}.}

\begin{tabular}{cc}

        \begin{minipage}[c]{0.46\linewidth}
            \centering
            \subcaption{Cluster Similarity}
            \label{comparisons:similarity}
\begin{tabular}{|c|c|c|}
\hline
& Panel Data  & Wearable Data \\ \hline 
VISTA   & 0.90   & 0.953    \\ \hline 
DTW     & 0.70 & 0.411 \\ \hline 
VAR     & 0.90 & 0.415 \\ \hline 
ResNet  & 0.90 & 0.458 \\ \hline 
DWT     & 0.90 & 0.335 \\ \hline
\end{tabular} 
        \end{minipage}

        &

        \begin{minipage}[c]{0.46\linewidth}
            \centering
            \subcaption{Adjusted Rand Index}
            \label{comparisons:ari}
\begin{tabular}{|c|c|c|}
\hline
& Panel Data  & Wearable Data \\ \hline 
VISTA   & 0.62    & 0.875     \\ \hline 
DTW     & 0.13 & 0.116 \\ \hline 
VAR     & 0.62 & 0.116 \\ \hline 
ResNet  & 0.62 & 0.094 \\ \hline 
DWT     & 0.62 & 0.041 \\ \hline
\end{tabular} 
        \end{minipage}
    \end{tabular}
\label{comparisons}
\end{table}

\subsection{Epidemiological Data: COVID-19} The COVID-19 dataset\footnote{\url{https://data.who.int/dashboards/covid19}, also available at \url{https://github.com/benjaminbrindle/vista_ssm} for easy access.} is comprised of weekly reports of the cumulative number of COVID-19 cases and deaths per 100,000 people from 53 European countries weekly from January 5, 2020 to May 3, 2023 (the last full week before the official pandemic end date of May 5, 2023 as established by the World Health Organization), totalling 174 reports. We transform the dataset to length 173 by instead considering the differences in number between each week, then dividing each feature by the maximum obtained by that feature across all countries and time points to ensure the data is between 0 and 1. 

We showcase VISTA's applicability in this case with no ground truth labels. For comparison, we reference \parencite{srivastava2023spatial}, in which authors employ an approach derived from functional data analysis to cluster the same European countries based on the cases feature alone from February to October 2020. The authors assign the COVID-19 time series to four clusters and demonstrate that each cluster is characterized by distinct temporal features such as the number and amplitude of waves of new infections.

\begin{table}[H]
\centering
    \caption{
        ABIC from identity initialization for varying $M$ and $d$.
    }
\label{covid:bic}
\begin{tabular}{|c| c| c| c| c|c|c|}
\hline
\diagbox{$d$}{$M$} & 3 & 4 & 5 & 6 & 7 & 8 \\ \hline
2 & -9.44E+04 & -9.75E+04 & -9.87E+04 & -9.89E+04 & -9.91E+04 & -1.00E+05  \\ \hline
3 & -9.77E+04 & -1.01E+05 & -1.02E+05 & -1.03E+05 & -1.02E+05 & -1.03E+05  \\ \hline
4 & -9.85E+04 & \textbf{-1.05E+05} & -1.04E+05 & -1.05E+05 & -1.03E+05 & -1.04E+05  \\ \hline
5 & -1.00E+05 & -9.61E+04 & -1.03E+05 & -1.05E+05 & -1.07E+05 & -1.07E+05  \\ \hline
6 & -9.63E+04 & -1.04E+05 & -1.04E+05 & -1.06E+05 & \textbf{-1.08E+05} & -1.07E+05  \\ \hline
7 & -1.01E+05 & -9.58E+04 & -1.02E+05 & -1.07E+05 & -1.02E+05 & -1.06E+05  \\ \hline
8 & -1.01E+05 & -9.82E+04 & -1.02E+05 & -1.07E+05 & -1.06E+05 & -1.06E+05  \\ \hline
\end{tabular} 
\end{table}

\begin{figure}[H]
    \centering
    \includegraphics[width=0.8\linewidth]{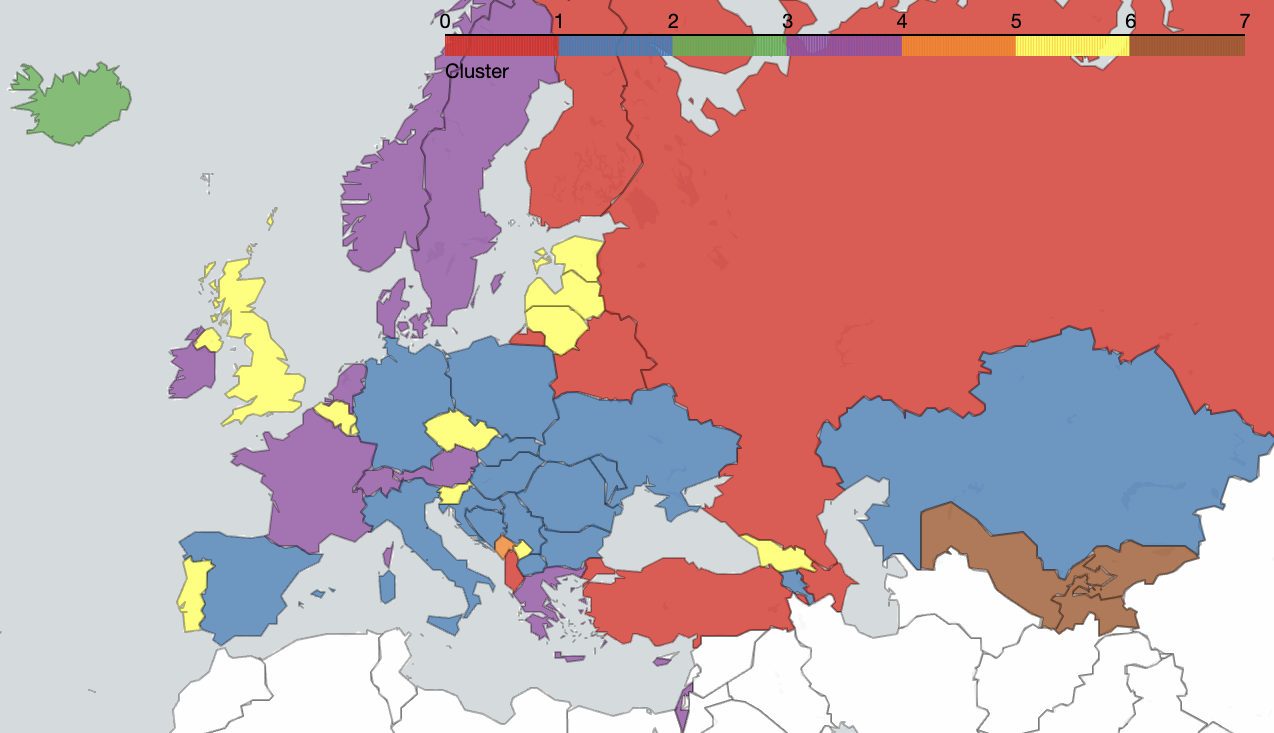}
    \caption{Color coding of countries by cluster membership for $d=6$ and $M=7$. \tiny{\textbf{Cluster 0:} Kyrgyzstan, Tajikistan, Uzbekistan; 
\textbf{Cluster 1:} Albania, Azerbaijan, Belarus, Finland, Russia, Turkey; 
\textbf{Cluster 2:} Armenia, Bosnia and Herz., Bulgaria, Croatia, Germany, Hungary, Italy, Kazakhstan, Macedonia, Poland, Moldova, Romania, Serbia, Slovakia, Spain, Ukraine; 
\textbf{Cluster 3:} Austria, Cyprus, Denmark, France, Greece, Ireland, Israel, Netherlands, Norway, Sweden, Switzerland; 
\textbf{Cluster 4:} Andorra, Monaco, Montenegro, San Marino; 
\textbf{Cluster 5:} Belgium, Czech Rep., Estonia, Georgia, Kosovo, Latvia, Lithuania, Luxembourg, Malta, Portugal, Slovenia, United Kingdom; 
\textbf{Cluster 6:} Iceland}}
    \label{fig:covid1}
\end{figure}

\begin{figure}[H]
    \centering
    \includegraphics[width=0.8\linewidth]{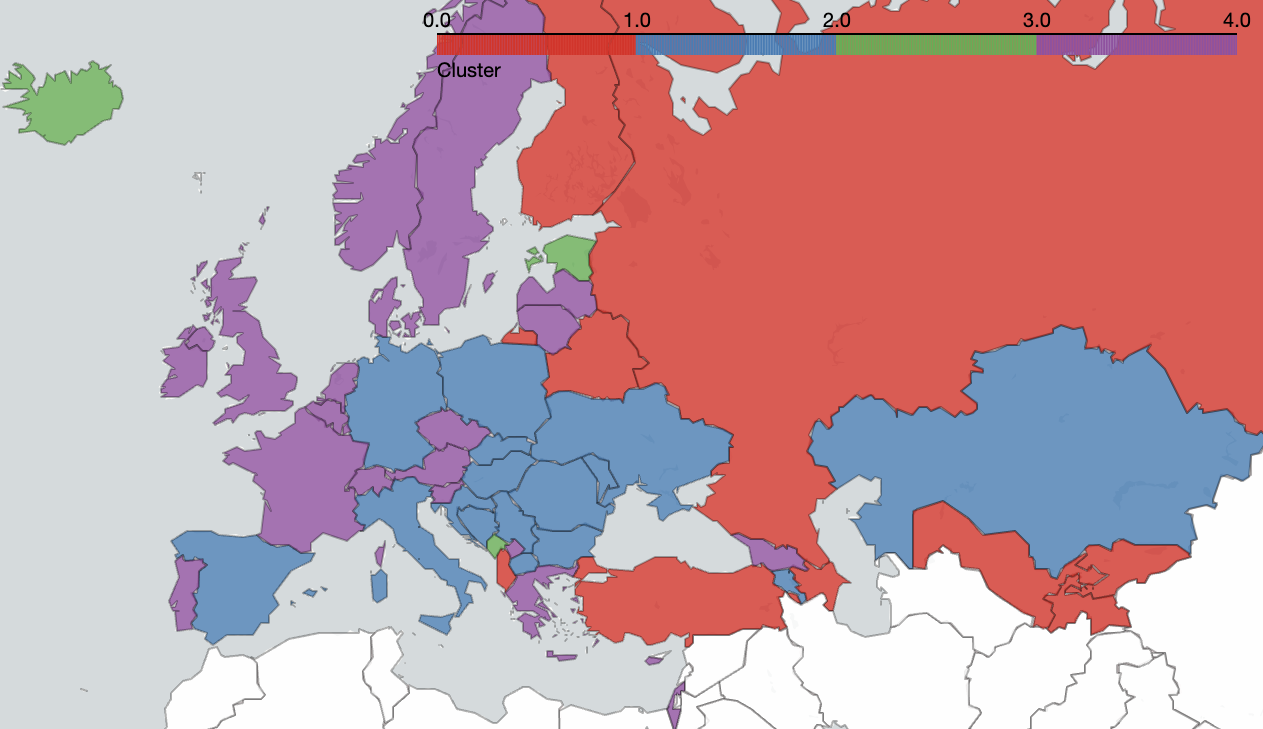}
    \caption{Color coding of countries by cluster membership for $d=4$ and $M=4$. \tiny{\textbf{Cluster 0:} Albania, Azerbaijan, Belarus, Finland, Kyrgyzstan, Russia, Tajikistan, Turkey, Uzbekistan; 
\textbf{Cluster 1:} Armenia, Bosnia and Herz., Bulgaria, Croatia, Germany, Hungary, Italy, Kazakhstan, Macedonia, Poland, Moldova, Romania, Serbia, Slovakia, Spain, Ukraine; 
\textbf{Cluster 2:} Andorra, Estonia, Iceland, Monaco, Montenegro, San Marino; 
\textbf{Cluster 3:} Austria, Belgium, Cyprus, Czech Rep., Denmark, France, Georgia, Greece, Ireland, Israel, Kosovo, Latvia, Lithuania, Luxembourg, Malta, Netherlands, Norway, Portugal, Slovenia, Sweden, Switzerland, United Kingdom}}
    \label{fig:covid2}
\end{figure}

We perform a similar task but combine both the number of new cases and new deaths into two-dimensional time series, and investigate the number of clusters in the dataset via the ABIC for model selection. We use the identity initialization for better reproducibility in this unlabeled setting. We begin by running VISTA on a range of values for $M$ and $d$ to determine the ABIC values shown in table \ref{covid:bic}. The lowest ABIC occurs when $d=6$ and $M=7$. In figure \ref{fig:covid1} we show the countries color coded on the map according to their cluster membership for the case of $d=6$ and $M=7$. In figure \ref{fig:covid2}, to compare with the results of \parencite{srivastava2023spatial}, we also display a map for the case of $d=4$ and $M=4$, which yields the lowest ABIC score in the four class setting. We do not expect to produce identical results to those of this earlier work as we add the deaths feature and more than two years of additional data. However, we notice that countries in eastern Europe are typically placed in a different cluster than those in western Europe, and that one of the clusters specifically groups countries with small populations and land area (e.g. Monaco, Andorra, San Marino...). Even more interesting is a comparison between figure \ref{fig:covid1} and figure \ref{fig:covid2}, where we observe some of the clusters in the $M=7$ case seeming to split from the clusters in the $M=4$ case, demonstrating some consistency of the clustering results despite different choices of the meta parameters $M$ and $d$.

\subsection{Ecological Momentary Assessment Data: Depression}

The ecological momentary assessments (EMA) dataset\footnote[1]{Collected from \url{https://osf.io/mvdpe/}} \parencite{fried2022mental} was created in March 2020 to capture mental health status, number of social interactions, and preoccupation with COVID-19 four times per day for 14 days for 80 undergraduate students at Leiden University in the Netherlands. Participants were issued 14-item EMA surveys at noon, 3pm, 6pm, and 9pm each day and had one hour to complete each survey after receiving it. Given the irregular number of surveys due to missing data and the variability in when individuals responded, this dataset provided a suitable empirical study featuring both irregularly sampled time series and time series of varying lengths. 

We selected 8 of the 14 EMA features to focus on items related to mental health \parencite{fried2022mental}: ``I found it difficult to relax'' (Relax), ``I was worried about different things'' (Worry), ``I couldn’t seem to experience any positive feeling at all'' (Anhedonia), ``I felt irritable'' (Irritable); ``I felt nervous, anxious, or on edge'' (Nervous); ``I felt that I had nothing to look forward'' (Future); ``I felt tired'' (Tired); and ``I felt like I lacked companionship'' (Alone). Each item records how much each respondent agrees with a statement about their mental health in a 1-5 likert scale, which we rescaled to be within the interval $[0,1]$. From the original 80 participants, we dropped those missing more than 40\% of their assessments (i.e., 23 or more out of 56), resulting in a final sample of 74 students. 

The original study identified an increase in average depressive symptoms in the sample, shifting from the normal to the mild depression range \parencite{fried2022mental}. Accordingly, we used EMA features to predict which participants would endorse mild to severe depressive symptoms by the end of the study as captured by the Depression Anxiety Stress Scale (DASS-21) \parencite{lovibond1995structure}. We used DASS-21 scores as ground truth for classification by grouping Normal symptoms severity in one category (DASS-21 Depression: score < 10) and Mild, Moderate, Severe, and Extremely Severe in another (score $\geq$ 10). The final ground-truth labels consisted of 37 subjects in the asymptomatic group and 37 in the symptomatic group.

Using the k-means initialization we evaluate the VISTA's performance on this dataset. In table \ref{mental:perf} we show ABIC values, determined by running VISTA 10 times for each value of $d$ and $M$ shown and averaging the ABIC for each. In this two-class setting, the lowest ABIC occurs when $d=14$. We show the cluster similarity and ARI in table \ref{mental:perf}. In table \ref{mental:cm} we display the confusion matrix of the execution of VISTA that returned the highest cluster similarity for $d=14$, which in this case is 0.770.

\begin{figure}[H]
    \centering
    \includegraphics[width=\linewidth]{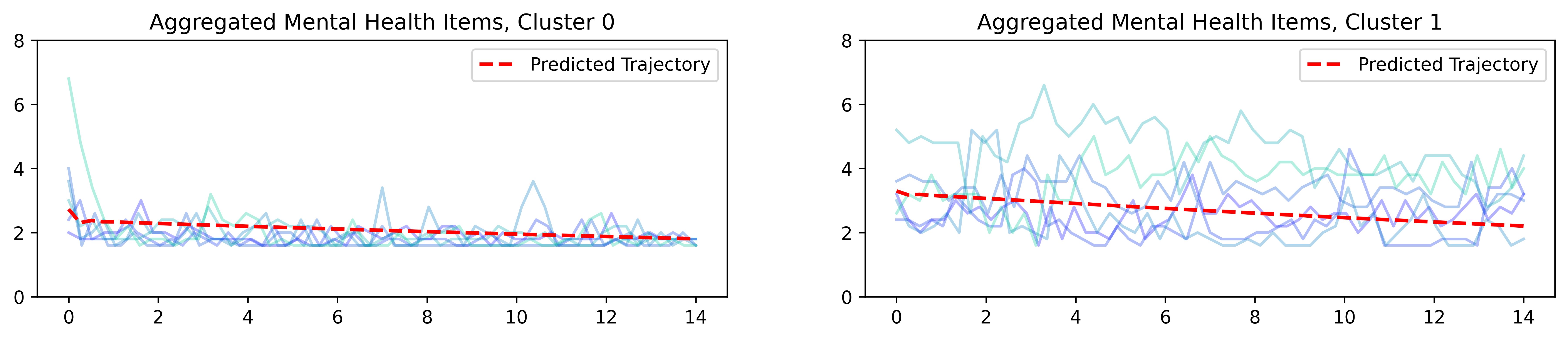}
    \caption{The result of VISTA with the returned clusters and noiseless trajectory determined by the optimal parameters $\mu$, $A$, and $C$ for the depression time series. The x-axis is in days, while the y-axis is the sum of all 8 mental health items used in clustering.}
    \label{fig:mental}
\end{figure}

In figure \ref{fig:mental}, we display a few of the time series that VISTA placed in each cluster, along with a noiseless trajectory predicted from the LGSSM parameters returned by VISTA with the same parameters returned from the case that generated the above confusion matrix. For both time series and trajectories, the plots are the result of summing over all 8 mental health features in the time series and generated LGSSM. Cluster 0 corresponds to the asymptomatic group, while cluster 1 corresponds to the symptomatic group.

\begin{table}[H]
    \caption{
        Mean (standard deviation) of ABIC, cluster similarity, and ARI from 10 trials with k-means initialization for $M=2$ clusters and fixed $d$ shown along with a confusion matrix for $d=14$ corresponding to the highest cluster similarity from the best average ABIC.
    }

    \begin{tabular}{c}
        
        \begin{minipage}[c]{\linewidth}
            \centering
            \subcaption{Performance metrics}
            \label{mental:perf}
            \begin{tabular}{|c|c|c|c|}
\hline
d  & ABIC                        & Similarity  & ARI         \\ \hline
8  & -4.22E+04 (1143.79)         & 0.74 (0.01) & 0.22 (0.02) \\ \hline
9  & -4.32E+04 (462.30)          & 0.74 (0.01) & 0.22 (0.02) \\ \hline
10 & -4.37E+04 (603.63) & 0.75 (0.01) & 0.24 (0.03) \\ \hline
11 & -4.39E+04 (556.05)          & 0.74 (0.01) & 0.22 (0.02) \\ \hline
12 & -4.37E+04 (591.24)          & 0.74 (0.02) & 0.23 (0.04) \\ \hline
13 & -4.40E+04 (712.57)          & 0.74 (0.01) & 0.22 (0.02) \\ \hline
14 & \textbf{-4.42E+04 (673.12) }         & 0.74 (0.01) & 0.22 (0.02) \\ \hline
15 & -4.32E+04 (1766.26)         & 0.72 (0.05) & 0.20 (0.06) \\ \hline
16 & -4.38E+04 (975.36)          & 0.74 (0.01) & 0.22 (0.02) \\ \hline
\end{tabular} 
            \vspace{5mm}
        \end{minipage}

\\
        \begin{minipage}[c]{\linewidth}
            \centering
            \subcaption{Best Confusion Matrix, $d=14$}
            \label{mental:cm}
            \begin{tabular}{|c|c|c|} \hline
                \diagbox{True}{Predicted} & Asymptomatic & Symptomatic \\ \hline
                Asymptomatic & 27 & 10 \\ \hline
                Symptomatic & 7 & 30 \\ \hline
            \end{tabular}
            \vspace{5mm}
        \end{minipage}

    \end{tabular}
    \label{mental_tables}
\end{table}

\section{Discussion}
We derived and proposed VISTA, an unsupervised approach for automatically identifying clusters in a cohort of time series without assuming regular or consistent sampling. VISTA is based on a generative parametric state space modeling of the observations as a mixture of LGSSMs with parameters fitted to the data via an expectation-maximization scheme (Figure \ref{fig:schematic}). We carefully adapted the stochastic dynamical system of LGSSM in order to account for varying and irregular sampling rates and developed a corresponding implementation that overcomes some shortcomings of past related frameworks. We validated and benchmarked VISTA against several time series clustering methods, and on a variety of datasets with different degrees of difficulty in terms of dimension, complexity and noise. 

Taken together, our initial evaluations on a diverse set of time series suggest that VISTA's ability to handle irregular temporal data could be leveraged in real-world applications, including psychological studies and digital health monitoring. These settings are typically characterized by unevenly distributed and noisy data collection \parencite{pratap2020indicators}, which can hinder accurate analysis \parencite{kreindler2016effects} and the detection of relevant subpopulations. As suggested by our classification findings, we anticipate that VISTA will be effective at modeling subgroups from ecological momentary assessments and wearable data, both of which are becoming increasingly common data collection sources in psychology and healthcare \parencite{galatzer2023machine}.

Another application of VISTA is modeling clinical language patterns through markers identified using language models and other natural language processing (NLP) methods \parencite{malg2023nlpreview}. Clinical corpora present unique modeling challenges due to their dynamic nature, causing NLP features from patients, therapists, and their dyadic conversations to vary considerably across time, individuals, and contexts. This variability makes it difficult to model meaningful clusters based on linguistic features and to associate these clusters with group-level outcomes \parencite{malgaroli2024linguistic}. These challenges become even more pronounced in asynchronous text-based interventions, where inconsistent communication frequencies and diverse messaging styles add further complexity to NLP analyses \parencite{hull2020two}. Applying VISTA in these settings could help identify subgroups based on heterogeneous NLP features captured from clinical text, while accounting for irregularities and variability in language. These clusters of NLP markers have the potential to capture emotional or psychological features, which could then be analyzed for their associations with key clinical indicators such as symptom severity, treatment adherence, and intervention outcomes.

Despite its strengths, there are several remaining limitations of this work that could form the basis for future investigations. First, the EM algorithm is particularly sensitive to changes in initialization, offering no guarantees of reaching a global maximum. We addressed this in our analyses by running VISTA multiple times, each with a different initialization. However, other initialization schemes beyond the three we provide could generate superior results for specific problems. Second, parameter identifiability also poses a challenge in this setting, as certain invariances in the model formulation exist. This limitation makes it infeasible to guarantee recovery of underlying parameters generating the data without imposing further constraints. To account for this, VISTA includes an option to fix parameters for specific use cases. In the case of a single LGSSM, such as those used in the k-means initialization, multiple choices of parameters can often be used to parametrize similar dynamics. Our mixture modeling approach, on the other hand, fits LGSSMs to multiple time series simultaneously, mitigating some identifiability concerns compared to single time series analysis. These points contribute to the overall stability of VISTA’s clustering results, which we observed in the consistency of clustering results in the epidemiological data example. Additionally, our assumption that all noise is Gaussian may not be appropriate for every application. A future extension of this work could be the incorporation of non-Gaussian noise. Furthermore, VISTA is unable to automatically determine the optimal number of clusters for a given dataset. We provide information criteria as a way to interpret the likelihood and determine the optimal number of clusters, as shown in the epidemiological data example. However, depending on individual use cases, domain-specific knowledge or additional exploratory analysis could be necessary to determine the fixed number of clusters with which to run VISTA. A direction for future work could be in automatically determining the optimal number of clusters, though this could prove computationally expensive on large datasets. Beyond these possible directions for future extensions, we also recognize the potential to extend this framework by incorporating a control variable on either the state or the observed variable. We envision adding a unique control to each cluster, providing the model with a deterministic trend per cluster to better capture complex temporal patterns.

These limitations notwithstanding, our findings suggest that VISTA is a flexible and powerful tool for modeling irregular time series. To promote transparency and reproducibility, we have made our Python code and detailed examples publicly available at \url{https://github.com/benjaminbrindle/vista_ssm}. By sharing these resources, we aim to empower researchers to leverage VISTA in tackling the unique challenges posed by healthcare and psychological data.

\section{Acknowledgments}
MM was supported by the National Institute of Mental Health (NIMH) award K23MH134068. NC was supported by the National Science Foundation (NSF), award 2402555. The content is solely the responsibility of the authors and does not necessarily represent the official views of the NIMH or NSF.

\section{Author Contributions}
Conceptualization: BB, TDH, NC, MM. Methodology: BB, MM, NC. Software: BB, MM, NC. Validation: BB, MM, NC. Formal analysis: BB, MM, NC. Investigation: BB, MM, NC. Resources:  BB, MM, NC. Data Curation:  BB, MM, NC. Writing  (Original Draft): MM, NC, BB, TDH. Writing (Review and Editing): MM, NC, BB, TDH. Visualization: BB, MM, NC. Supervision: MM, NC. Project administration:  MM, TDH. Funding acquisition: MM, NC, TDH.  All authors reviewed the results and approved the final version of the manuscript.

\nocite{*}
\printbibliography

\end{document}